\documentclass[aps,prl,twocolumn,groupedaddress]{revtex4-1}
\usepackage{graphicx}
\usepackage{amsmath}
\usepackage{amssymb}
\usepackage{braket}
\usepackage{bm}

\begin{document}

\title{Fitness response relation of a multi-type age-structured population dynamics}
\author{Yuki Sughiyam$\mathrm{a}^{1}$, So Nakashim$\mathrm{a}^{2}$ and Tetsuya J. Kobayash$\mathrm{i}^{1,3}$}
\affiliation{${}^{1}\mathrm{I}$nstitute of Industrial Science, The University of Tokyo, 4-6-1, Komaba, Meguro-ku, Tokyo 153-8505 Japan}
\affiliation{${}^{2}\mathrm{G}$raduate School of Information and Technology, Department of Mathematical Informatics, The University of Tokyo, 7-3-1, Hongo, Bunkyo-ku, Tokyo 113-8654 Japan}
\affiliation{${}^{3}\mathrm{P}$RESTO, Japan Science and Technology Agency (JST), 4-1-8, Honcho, Kawaguchi, Saitama 332-0012 Japan}
\date{\today}
\begin{abstract}
We construct a pathwise formulation for a multi-type age-structured population dynamics, which involves an age-dependent cell replication and transition of gene- or phenotypes. 
By employing the formulation, we derive a variational representation of the stationary population growth rate; the representation comprises a trade-off relation between growth effects and a single-cell intrinsic dynamics described by a semi-Markov process. 
This variational representation leads to a response relation of the stationary population growth rate, in which statistics on a retrospective history work as the response coefficients. 
These results contribute to predicting and controlling growing populations based on experimentally observed cell-lineage information. 
\end{abstract}
\maketitle
\section{I. Introduction}
Predicting and controlling evolution rather than reconstructing it has been one of the pivotal challenges in evolutionary biology \cite{a3}. 
Because the natural selection favors changes in a population that can increase the fitness of the population, predicting fitness response to perturbations is crucial for anticipating the future evolutionary path. 
In the case of controlling evolution, in contrast, the fitness response relation can guide us in how to reduce fitness rather than increasing it to suppress the outbreak of malignant pathogens and tumors by drug applications or specific treatments \cite{01}. 

The evaluation of the fitness or its proxy, the population growth rate, has been conducted in the context of ordinary or partial differential equations by focusing on the time-slice distribution of the population \cite{06,07,08,09}. 
In these approaches, the problems are mostly reduced to eigenvalue problems of the differential equations with appropriate boundary conditions, the largest eigenvalue of which corresponds to the stationary population growth rate. 
However, an eigenvalue is typically a complicated or an implicit function of underlying parameters of the population dynamics. 
As a result, the direct evaluation of the fitness response from differential equations is not practical especially when we apply the result to experimental data. 

The recent introduction of a pathwise formulation of the population dynamics \cite{11,12,14,15,16,17} has partially resolved this discrepancy between the theory and the experiments by revealing a variational representation of the population growth rate and the associated fitness response relation. 
The formulation was motivated for analyzing cell-lineage data obtained by novel experimental technologies to trace replicating cells over hundreds of generations under microfluidic devices such as the mother machine \cite{10} and the dynamics cytometer \cite{11,12}. 
By using the pathwise formulation, Wakamoto {\it et al}. revealed a fitness response relation of an age-structured but phenotypically homogeneous population \cite{11}. 
Another fitness response relation was also obtained for phenotypically heterogeneous but not age-structured population \cite{16,17}. 
In both relations, the sensitivities of the fitness to perturbations were characterized by empirical age or type distributions evaluated over a sufficiently long genealogical path of a survived cell, which is directly measurable by the long-term cell tracing experiments. 
This shared properties strongly imply the generality and universality of the response relations. 
Nevertheless, neither of them is sufficiently practical for applying to actual populations of cells, each member of which replicates and dies depending on both its age and phenotypic state. 

In this work, we unify these two complementary lines of studies by deriving a variational structure of a multi-type age-structured population dynamics (MTASP) and its associated response relation. 
This paper is organized as follows. 
In the next section, before working on the MTASP, we introduce a single-cell dynamics by ignoring birth and death effects such as a cell death and sister cells generated by cell division. 
Here, we show that a history (path) of the single-cell dynamics, which is observed in the mother machine, can be described by a semi-Markov process \cite{18,19,20}. 
In Sec. III, by incorporating the birth and death effects, we formulate the MTASP, which is represented by a McKendric equation with the type transition \cite{06,07,08,09}. 
In Sec. IV, we show a pathwise representation of the MTASP. 
Here, it is revealed that the time-backward (retrospective) path probability observed in the dynamics cytometer is given by biasing the time-forward (chronological) path probability describing the single-cell dynamics. 
In Sec. V, by employing the large deviation theory \cite{21}, we derive a coarse-graining picture of the relationship between time-forward and time-backward paths. 
We also find that the difference between a time-forward and time-backward rate functions gives the stationary population growth rate. 
In Sec. VI and VII, we show that the variational principle relates the stationary population growth rate with the time-forward rate function via the Legendre transformation. 
By using this relation, we also derive the response relation for the population growth rate, which shows that the response coefficients can be evaluated by some statistics on the time-backward path. 
In Sec. VIII, we derive the explicit form of the time-backward rate function and show that the time-backward path can be mimicked by a biased semi-Markov process called a retrospective process. 
In Sec. IX, we verify the analytical results derived in the previous sections by numerical simulations. 
Finally, we summarize this paper in Sec. X. 

\section{II. Single-cell dynamics of age and type}
We begin with describing the single-cell dynamics that can be observed by tracing a dividing cell by ignoring its sister cells generated by cell divisions. 
Such dynamics is measured by using the mother machine \cite{10} as in FIG. 1 (A). 
\begin{figure}[h]
\begin{center}
\includegraphics*[height=10cm]{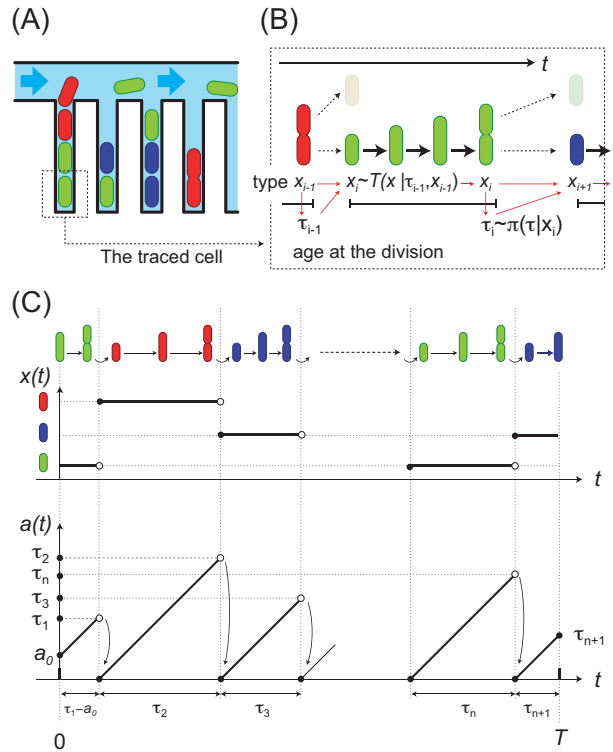}
\caption{(A) A schematic illustration of single-cell tracing by the mother machine. 
(B) A schematic diagram of the stochastic laws of the single-cell dynamics. 
The inter-division interval and the type transition are governed by $\pi\left(\tau|x\right)$ and $\mathbb{T}\left(x|\tau^{\prime},x^{\prime}\right)$, respectively. 
(C) A schematic illustrations of the transition dynamics of type $x\left(t\right)$ (the upper panel) and the age $a\left(t\right)$ (the lower panel). 
Upon a division event, the type transition occurs and the age resets to $0$. 
Also, we denote an age of a mother cell at $i$th division event by $\tau_{i}$. }
\end{center}
\end{figure}
Let $a\in\left[0,\infty\right)$ and $ x\in\Omega$ be an age and a type of cells, respectively, where $\Omega$ denotes a finite state set of $x$. 
The age $a$ is defined as the elapsed time since the last division. 
In this section, we tentatively assume that the cell is immortal and can be traced for infinitely long time. 

In the time evolution, the cell divides with an age- and type-dependent rate $r\left(a,x\right)\geq 0$. 
With this rate, the probability density function of the inter-division interval can be represented as 
\begin{equation}
\pi\left(\tau|x\right):=r\left(\tau,x\right)e^{-\int_{0}^{\tau}r\left(a,x\right)da},\label{defpi}
\end{equation}
where $e^{-\int_{0}^{\tau}r\left(a,x\right)da}$ is the probability that the cell did not divide up to time $\tau$ 
and $ r\left(\tau,x\right)d\tau$ is the probability that the cell commits division at age $\tau$ (see FIG. 1 (B)). 

Next, we denote the probability of the type transition from $x^{\prime}$ to $x$ upon division by $\mathbb{T}\left(x|\tau^{\prime},x^{\prime}\right)$, 
where the transition is supposed to be dependent on the age of the cells at the division, $\tau^{\prime}$. 
In this work, we suppose that the type represents either genotype or epigenetic state that can change only upon division by mutation or error in epigenetic state transmission, respectively. 
Also, we assume that the transition matrix $\mathbb{T}\left(x|\tau^{\prime},x^{\prime}\right)$ is primitive (ergodic) for any age $\tau^{\prime}$: 
for any age $\tau^{\prime}$, there exists a certain natural number $m$ such that $\mathbb{T}^{m}\left(x_{m}|\tau^{\prime},x_{0}\right):=\Sigma_{\left\{x_{i}\right\}_{i=1}^{m-1}}\Pi_{i=1}^{m}\mathbb{T}\left(x_{i}|\tau^{\prime},x_{i-1}\right)>0$ for all pairs of indices $ x_{m},x_{0}\in\Omega$. 
This assumption can be biologically rephrased as follows: 
a cell can aperiodically reach any types by the finite number of transitions. 
Then, the joint dynamics of both division and type can be described by a semi-Markov process \cite{18,19,20} generated by the semi-Markov kernel: 
\begin{equation}
Q\left(x;\tau^{\prime}|x^{\prime}\right):=\mathbb{T}\left(x|\tau^{\prime},x^{\prime}\right)\pi\left(\tau^{\prime}|x^{\prime}\right).\label{defQ}
\end{equation}

Suppose that we observe a cell up to time $T$. 
Then, the history (path) of the cell, $\chi_{T}:=\left\{n;a_{0},x_{1},\tau_{1},x_{2},\tau_{2},...,x_{n},\tau_{n},x_{n+1},\tau_{n+1}\right\}$, can be characterized 
by the initial age of the cell $a_{0}$, the number of division events $n$ up to $T$, the inter-division intervals $\left\{\tau_{i}\right\}_{i=1}^{n+1}$, and the types $\left\{x_{i}\right\}_{i=1}^{n+1}$ before the $i$th division (see FIG. 1 (C)). 
For notational simplicity, $\tau_{n+1}$ is specially defined as the time interval between $n$th division and $T$. 
Also, since $\tau_{1}$ represents the inter-division interval until the $1$st division, the time interval between the start of the observation and the $1$st division is calculated as $\tau_{1}-a_{0}$ (see FIG. 1 (C)). 
The probability density $\mathbb{P}_{F}\left[\chi_{T}\right]$ to observe such a history is obtained by multiplying the semi-Markov kernel over the history as
\begin{eqnarray}
\nonumber\mathbb{P}_{F}\left[\chi_{T}\right]&=&\delta\left(T-\left\{\sum_{i=1}^{n+1}\tau_{i}-a_{0}\right\}\right)\times e^{-\int_{0}^{\tau_{n+1}}r\left(a,x_{n+1}\right)da}\\
\nonumber&&\times\left\{\prod_{i=1}^{n}\mathbb{T}\left(x_{i+1}|\tau_{i},x_{i}\right)\pi\left(\tau_{i}|x_{i}\right)\right\}\\
&&\times e^{\int_{0}^{a_{0}}r\left(a,x_{1}\right)da}\rho_{0}\left(a_{0},x_{1}\right),\label{PrF}
\end{eqnarray}
where $\rho_{0}\left(a_{0},x_{1}\right)$ denotes the probability density to sample a cell with age $a_{0}$ and type $x_{1}$ at the start of the observation. 
Also, we note that we used the fact that the probability until the $1$st division can be represented as 
$r\left(\tau_{1},x_{1}\right)e^{-\int_{a_{0}}^{\tau_{1}}r\left(a,x_{1}\right)da}\rho_{0}\left(a_{0},x_{1}\right)=\pi\left(\tau_{1}|x_{1}\right)e^{\int_{0}^{a_{0}}r\left(a,x_{1}\right)da}\rho_{0}\left(a_{0},x_{1}\right)$. 
While $\mathbb{P}_{F}\left[\chi_{T}\right]$ characterizes the intrinsic division and type transition dynamics of the single cell, 
$\chi_{T}$ may not typically be observed in a growing population of cells, 
because $\mathbb{P}_{F}\left[\chi_{T}\right]$ is obtained by ignoring the death of the cell and the fates of the sister cells. 
If a cell divides more frequently than others, the cell has more daughter cells than the others, which means that its tribe is overrepresented in the population. 
Moreover, by considering death of the cell, a certain history $\chi_{T}$ may be less observable than being expected from $\mathbb{P}_{F}\left[\chi_{T}\right]$ 
if a cell with the history $\chi_{T}$ is more likely to die than cells with other histories. 

\section{III. Multi-type age-structured population dynamics}
In order to appropriately account for the contribution of the sister cells and cell death that have been ignored in the previous section, 
we introduce a model of a multi-type age-structured population dynamics (MTASP) (see FIG. 2). 
Such dynamics is observed by using the dynamics cytometer \cite{11,12} as in FIG. 2 (A). 
\begin{figure}[h]
\begin{center}
\includegraphics*[height=9cm]{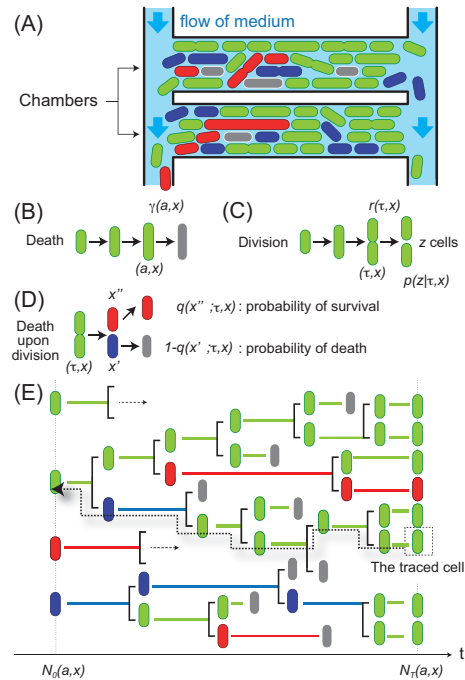}
\caption{(A) A schematic illustration of the dynamics cytometer. 
(B) (C) (D) A schematic images of the cell death, division and death upon division, respectively.
(E) A schematic illustration of the lineage tree generated by the MTASP. 
Such a lineage tree can be observed by the the dynamics cytometer. 
The gray color denotes cell death. 
The dashed arrow represents a retrospective tracking.}
\end{center}
\end{figure}

Let $\gamma\left(a,x\right)$ be the rate of cell death, which is dependent on both age $a$ and type $x$ of a cell. 
Then, the probability that a cell can survive from age $0$ up to age $\tau$ can be represented as $e^{-\int_{0}^{\tau}\gamma\left(a,x\right)da}$ (see FIG. 2 (B)). 
Next, suppose that a cell that divides at age $\tau^{\prime}$ and with type $x^{\prime}$ asexually generates $z^{\prime}$ number of its descendants including itself 
with a probability $p\left(z^{\prime}|\tau^{\prime},x^{\prime}\right)$ (see FIG. 2 (C)). 
Because the age of the divided cells is reseted to $0$ after division, all $z^{\prime}\geq 2$ descendants are equivalent. 
The number of descendants $z^{\prime}$ is typically $2$, but can be more than $2$ 
when the cell is filamentous with multiple chromosomal copies, which can generate multiple sister cells simultaneously \cite{a2}. 
Upon division, each daughter cell stochastically changes its type to $x$ depending on $\mathbb{T}\left(x|\tau^{\prime},x^{\prime}\right)$ (see FIG. 1 (B)). 
During and upon the transition, the cell is assumed to experience a mortality risk and 
can survive with the probability $q\left(x;\tau^{\prime},x^{\prime}\right)$ depending on the age of the mother cell at the division, $\tau^{\prime}$, and the types $\left(x,x^{\prime}\right)$ before and after the division (see FIG. 2 (D)). 
Thus, the cell dies with the probability $1-q\left(x;\tau^{\prime},x^{\prime}\right)$ due to the type transition. 
$q\left(x;\tau^{\prime},x^{\prime}\right)$ can be regarded as the death induced by division-related intracellular events or by deleterious mutations. 

Under this setup, the time evolution of the expected number of cells with age $a$ and type $x$, which we denote by $N_{t}\left(a,x\right)$, 
can be described by the McKendric equation \cite{06,07,08,09}: 
\begin{equation}
\displaystyle \frac{\partial}{\partial t}N_{t}\left(a,x\right)=\left[-\frac{\partial}{\partial a}-\left\{\gamma\left(a,x\right)+r\left(a,x\right)\right\}\right]N_{t}\left(a,x\right),\label{McPD}
\end{equation}
with a boundary condition, 
\begin{eqnarray}
\displaystyle \nonumber N_{t}\left(0,x\right)&=&\displaystyle \sum_{x^{\prime}\in\Omega}\int_{0}^{\infty}d\tau^{\prime}\,q\left(x;\tau^{\prime},x^{\prime}\right)\mathbb{T}\left(x|\tau^{\prime},x^{\prime}\right)\\
&&\times b\left(\tau^{\prime},x^{\prime}\right)r\left(\tau^{\prime},x^{\prime}\right)N_{t}\left(\tau^{\prime},x^{\prime}\right),\label{bMcPD}
\end{eqnarray}
where $b\left(\tau^{\prime},x^{\prime}\right)$ represents the expected number of the daughter cells: $b\left(\tau^{\prime},x^{\prime}\right):=\Sigma_{z^{\prime}=1}^{\infty}z^{\prime}p\left(z^{\prime}|\tau^{\prime},x^{\prime}\right)$. 
Here, the terms, $\partial/\partial a,\ \gamma\left(a,x\right)$ and $r\left(a,x\right)$ in Eq. (\ref{McPD}) represent the effects by aging, death and division, respectively; 
also, the boundary condition Eq. (\ref{bMcPD}) can be interpreted 
such that the daughter cells  rejoin to the time evolution Eq.(\ref{McPD}) as neonates $N_{t}\left(0,x\right)$. 
A derivation of this equation (\ref{McPD}) is shown in Appendix A. 
Figure 2 (E) is a schematic illustration of the cell lineage tree representing the MTASP. 

The averaged expansion rate of the total population size (the stationary population growth rate) is mathematically defined as 
\begin{equation}
\displaystyle \lambda:=\lim_{t\rightarrow\infty}\frac{1}{t}\log\frac{N_{t}^{\mathrm{t}\mathrm{o}\mathrm{t}}}{N_{0}^{\mathrm{t}\mathrm{o}\mathrm{t}}}=\lim_{t\rightarrow\infty}\frac{1}{t}\log\frac{\sum_{x\in\Omega}\int_{0}^{\infty}da\,N_{t}\left(a,x\right)}{\sum_{x\in\Omega}\int_{0}^{\infty}da\,N_{0}\left(a,x\right)},\label{SPGR}
\end{equation}
where the denominator $N_{0}^{\mathrm{t}\mathrm{o}\mathrm{t}}$ and the numerator $N_{t}^{\mathrm{t}\mathrm{o}\mathrm{t}}$ represent the total populations at time $0$ and time $t$, respectively. 
In the partial-differential-equation approach that is often employed in the field of mathematical demography \cite{06,07,08,09}, this growth rate is calculated by the largest eigenvalue of the operator in Eq. (\ref{McPD}): 
\begin{equation}
\displaystyle \hat{H}:=-\frac{\partial}{\partial a}-\left\{\gamma\left(a,x\right)+r\left(a,x\right)\right\},\label{H}
\end{equation}
under the boundary condition Eq. (\ref{bMcPD}). 
(Detailed calculation to obtain $\lambda$ in this approach is shown in Appendix B.) 
In contrast, we employ a path-wise formulation in this study, 
in order to relate the single-cell dynamics $\mathbb{P}_{F}\left[\chi_{T}\right]$ with the population-level quantities such as $\lambda$. 
In the course of introducing the formulation, a retrospective picture hidden in the MTASP will be revealed. 

\section{IV. Pathwise representation of the multi-type age-structured population dynamics}
Let $N_{T}\left[\chi_{T}\right]$ be the expected number of the cells at time $T$ that have a history (path) $\chi_{T}=\left\{n;a_{0},x_{1},\tau_{1},x_{2},\tau_{2},...,x_{n},\tau_{n},x_{n+1},\tau_{n+1}\right\}$. 
We can evaluate $N_{T}\left[\chi_{T}\right]$ as
\begin{eqnarray}
\nonumber N_{T}\left[\chi_{T}\right]&=&\delta\left(T-\left\{\sum_{i=1}^{n+1}\tau_{i}-a_{0}\right\}\right)\\
\nonumber&&\times e^{-\int_{0}^{\tau_{n+1}}\left\{\gamma\left(a,x_{n+1}\right)+r\left(a,x_{n+1}\right)\right\}da}\\
\displaystyle \nonumber&&\times\biggl[\prod_{i=1}^{n}q\left(x_{i+1};\tau_{i},x_{i}\right)\mathbb{T}\left(x_{i+1}|\tau_{i},x_{i}\right)\\
\nonumber&&\times b\left(\tau_{i},x_{i}\right)e^{-\int_{0}^{\tau_{i}}\gamma\left(a,x_{i}\right)da}\pi\left(\tau_{i}|x_{i}\right)\biggr]\\
&&\times e^{\int_{0}^{a_{0}}\left\{\gamma\left(a,x_{1}\right)+r\left(a,x_{1}\right)\right\}da}N_{0}\left(a_{0},x_{1}\right),\label{Nagai}
\end{eqnarray}
where $N_{0}\left(a_{0},x_{1}\right)$ denotes the number of the cells having the age $a_{0}$ and type $x_{1}$ at time $0$. 
For $n=0$ (no division cases), the product part in Eq. (\ref{Nagai}) vanishes (we define $\Pi_{i=1}^{0}=1$), 
and we obtain $N_{T}\left[\left\{0;a_{0},x_{1},\tau_{1}\right\}\right]=\exp\left[-\int_{a_{0}}^{a_{0}+T}\left\{\gamma\left(a,x_{1}\right)+r\left(a,x_{1}\right)\right\}da\right]N_{0}\left(a_{0},x_{1}\right)$. 
The details of a derivation of Eq. (\ref{Nagai}) are shown in Appendix C. 
Here, we note that Eq. (\ref{Nagai}) consists of iterations of the following kernel except the initial and final parts: 
\begin{eqnarray}
\nonumber&& q\left(x_{i+1};\tau_{i},x_{i}\right)\mathbb{T}\left(x_{i+1}|\tau_{i},x_{i}\right)\\
&&\times b\left(\tau_{i},x_{i}\right)e^{-\int_{0}^{\tau_{i}}\gamma\left(a,x_{i}\right)da}\pi\left(\tau_{i}|x_{i}\right).\label{kernel}
\end{eqnarray}
The first line, $q\left(x_{i+1};\tau_{i},x_{i}\right)\mathbb{T}\left(x_{i+1}|\tau_{i},x_{i}\right)$, represents the fraction of the daughter cells that had switched their type from $x_{i}$ to $x_{i+1}$ 
and survived after the transition. 
On the other hand, the second line, $b\left(\tau_{i},x_{i}\right)e^{-\int_{0}^{\tau_{i}}\gamma\left(a,x_{i}\right)da}\pi\left(\tau_{i}|x_{i}\right)$, is the expected number of the cells generated by a mother cell with the type $x_{i}$ that has survived until the age $\tau_{i}$ and has committed cell division at the age $\tau_{i}$. 
By using Eq. (\ref{PrF}) and assuming that the initial population can be written as $\rho_{0}\left(a_{0},x_{1}\right)=N_{0}\left(a_{0},x_{1}\right)/N_{0}^{\mathrm{t}\mathrm{o}\mathrm{t}}$, we obtain
\begin{eqnarray}
\nonumber&& N_{T}\left[\chi_{T}\right]=\\
\nonumber&& e^{-\int_{0}^{\tau_{n+1}}\gamma\left(a,x_{n+1}\right)da}e^{\sum_{i=1}^{n}k\left(x_{i+1}:\tau_{i},x_{i}\right)}e^{\int_{0}^{a_{0}}\gamma\left(a,x_{1}\right)da}\\
&&\times \mathbb{P}_{F}\left[\chi_{T}\right]N_{0}^{\mathrm{t}\mathrm{o}\mathrm{t}},\label{Nhist}
\end{eqnarray}
where we define the growth kernel $k\left(x_{i+1}:\tau_{i},x_{i}\right)$ as 
\begin{equation}
e^{k\left(x_{i+1}:\tau_{i},x_{i}\right)}:=q\left(x_{i+1};\tau_{i},x_{i}\right)b\left(\tau_{i},x_{i}\right)e^{-\int_{0}^{\tau_{i}}\gamma\left(a,x_{i}\right)da}.\label{defk}
\end{equation}
The second line in Eq. (\ref{Nhist}) summarizes the effect of growth and death, whereas the third line expresses the single-cell dynamics. 
Because of this specific contribution of the second line, we define a pathwise growth kernel $\mathbb{K}\left[\chi_{T}\right]$ by the second line: 
\begin{eqnarray}
\nonumber&&\mathbb{K}\left[\chi_{T}\right]:=\\
&&e^{-\int_{0}^{\tau_{n+1}}\gamma\left(a,x_{n+1}\right)da}e^{\sum_{i=1}^{n}k\left(x_{i+1}:\tau_{i},x_{i}\right)}e^{\int_{0}^{a_{0}}\gamma\left(a,x_{1}\right)da}.\label{Kpathchi}
\end{eqnarray}

Because $N_{T}\left[\chi_{T}\right]$ is the number of the cells at time $T$ that have the history $\chi_{T}$, 
the probability to observe a history $\chi_{T}$ by retrospectively tracking a randomly sampled cell at time $T$ \cite{04,05,11,16,17,a9} (see FIG. 2 (E)) can be evaluated by normalizing $N_{T}\left[\chi_{T}\right]$ as 
\begin{eqnarray}
\displaystyle \nonumber\mathbb{P}_{B}\left[\chi_{T}\right]&:=&\displaystyle \frac{N_{T}\left[\chi_{T}\right]}{N_{T}^{\mathrm{t}\mathrm{o}\mathrm{t}}}=\frac{e^{\mathbb{K}\left[\chi_{T}\right]}\mathbb{P}_{F}\left[\chi_{T}\right]}{\left\langle e^{\mathbb{K}\left[\chi_{T}\right]}\right\rangle_{\mathbb{P}_{F}\left[\chi_{T}\right]}}\\
&=&e^{\mathbb{K}\left[\chi_{T}\right]-\Lambda_{T}}\mathbb{P}_{F}\left[\chi_{T}\right],\label{Pbpath}
\end{eqnarray}
where $\left\langle\cdot\right\rangle_{\mathbb{P}_{F}\left[\chi_{T}\right]}$ denotes the average over all paths during the time interval $\left[0,T\right]$ with respect to $\mathbb{P}_{F}\left[\chi_{T}\right]$. 
Also, $\Lambda_{T}$ is the population growth during time interval $\left[0,T\right]$, which is defined by 
\begin{equation}
\displaystyle \Lambda_{T}:=\log\frac{N_{T}^{\mathrm{t}\mathrm{o}\mathrm{t}}}{N_{0}^{\mathrm{t}\mathrm{o}\mathrm{t}}}=\log\left\langle e^{\mathbb{K}\left[\chi_{T}\right]}\right\rangle_{\mathbb{P}_{F}\left[\chi_{T}\right]}.\label{Blam}
\end{equation}
We should note that the number of division events, $n$, is stochastic, therefore we need to integrate it out in the average: 
\begin{equation}
\displaystyle \left\langle\cdot\right\rangle_{\mathbb{P}_{F}\left[\chi_{T}\right]}:=\sum_{n=0}^{\infty}\sum_{\left\{x_{i}\right\}_{i=0}^{n}}\int_{0}^{\infty}\cdots\int_{0}^{\infty}da_{0}\prod_{i=1}^{n+1}d\tau_{i}\left[\cdot\right]\mathbb{P}_{F}\left[\chi_{T}\right].
\end{equation}
The deviation of $\mathbb{P}_{B}\left[\chi_{T}\right]$ from $\mathbb{P}_{F}\left[\chi_{T}\right]$ clarifies that the chance to observe a certain history of a cell $\chi_{T}$ differs, 
depending on how we sample and trace a cell. 
A history $\chi_{T}$ is observed with the probability density $\mathbb{P}_{F}\left[\chi_{T}\right]$ when we sample a cell at the beginning of an experiment 
and trace it chronologically over time by ignoring its sister cells as in the mother machine (see FIG. 1 (A, B)). 
In contrast, $\chi_{T}$ is observed with the probability density $\mathbb{P}_{B}\left[\chi_{T}\right]$ when we sample a cell at the end of an experiment from a cultured population 
and trace its ancestors back in a retrospective manner as in the dynamics cytometer (see FIG. 2 (A, E)). 
The difference between them is due to the fact that we have more chance to sample the histories that are overrepresented by faster divisions and lower death. 
In terms of stochastic processes, $\chi_{T}\sim \mathbb{P}_{B}\left[\chi_{T}\right]$ can be regarded as an exponentially biased process of the semi-Markov process $\chi_{T}\sim \mathbb{P}_{F}\left[\chi_{T}\right]$ by the growth kernel $\mathbb{K}\left[\chi_{T}\right]$. 
We note that Eqs. (\ref{Pbpath}) and (\ref{Blam}) are formally the same as the correspondence between the chronological and the retrospective paths employed in Refs. \cite{04,05,11,16,17,a9} under simpler and thereby less realistic models. 
This means that this correspondence is a quit general structure underlaying the population dynamics. 

\section{V. Coarse-graining by contraction}
The pathwise representations of Eqs (\ref{PrF}) and (\ref{Pbpath}) hold for general situations. 
However, the potential variety of the possible paths is extremely huge, and therefore the path probabilities are not practically observed by experiments. 
In order to moderate the complexity and information that the pathwise representations have, 
we coarse grain a history $\chi_{T}$ with the following empirical distribution of triplets, $\left(x:\tau^{\prime},x^{\prime}\right)$, in the history $\chi_{T}$: 
\begin{equation}
j_{\mathrm{e}}\displaystyle \left(x;\tau^{\prime},x^{\prime}\right):=\frac{1}{T}\sum_{i=1}^{n}\delta_{x,x_{i+1}}\delta\left(\tau^{\prime}-\tau_{i}\right)\delta_{x^{\prime},x_{i}}.
\end{equation}
This empirical distribution measures how many times a division event with a type transition from $x^{\prime}$ to $x$ occurs 
at age $\tau^{\prime}$ in the history $\chi_{T}=\left\{n;a_{0},x_{1},\tau_{1},x_{2},\tau_{2},...,x_{n},\tau_{n},x_{n+1},\tau_{n+1}\right\}$ \cite{18}. 
Note that this empirical triplet depends on the history $\chi_{T}$, but we abbreviate it from the notation of $j_{\mathrm{e}}\left(x;\tau^{\prime},x^{\prime}\right)$ for simplicity. 
Also, we assume that the empirical distribution $j_{\mathrm{e}}\left(x;\tau^{\prime},x^{\prime}\right)$ is normalized as $\Sigma_{x,x^{\prime}\in\Omega} \int_{0}^{\infty} d\tau^{\prime}\,\tau^{\prime}j_{\mathrm{e}}\left(x;\tau^{\prime},x^{\prime}\right)=1$ for $ T\rightarrow\infty$. 
If paths are generated by following a path probability $\mathbb{P}\left[\chi_{T}\right]$, the probability to observe a certain $j\left(x;\tau^{\prime},x^{\prime}\right)$ is represented as 
$\mathbb{P}\left[j\right]:=\left\langle\mathfrak{I}\left[j_{\mathrm{e}}=j\right]\right\rangle_{\mathbb{P}\left[\chi_{T}\right]}$ where $\mathfrak{I}\left[j_{\mathrm{e}}=j\right]$ denotes the indicator functional: if $j_{\mathrm{e}}=j$, then $\mathfrak{I}\left[j_{\mathrm{e}}=j\right]=1$, otherwise $\mathfrak{I}\left[j_{\mathrm{e}}=j\right]=0$. 
For $ T\rightarrow\infty$, the empirical triplet $j_{\mathrm{e}}\left(x;\tau^{\prime},x^{\prime}\right)$ converges to the typical triplet $j^{*}\left(x;\tau^{\prime},x^{\prime}\right)$ due to the law of large numbers. 
If, for large but not infinite $T$, the probability to observe $j$ deviated from $j^{*}$ decays exponentially at rate $I\left[j\right]\geq 0$ 
and thereby $\mathbb{P}\left[j\right]$ is represented as $\mathbb{P}\left[j\right]\approx e^{-TI\left[j\right]},\ \chi_{T}\sim \mathbb{P}\left[\chi_{T}\right]$ is said to satisfy the large deviation principle and $I\left[j\right]$ is the rate function (also known as the large deviation function) \cite{21}. 
In the large deviation theory, the typical triplet $j^{*}\left(x;\tau^{\prime},x^{\prime}\right)$ is given by the argument attaining the minimum of the rate function (that is $0$), 
$I\left[j^{*}\right]=0$. 

The empirical triplet for the semi-Markov dynamics of a single cell, $\mathbb{P}_{F}\left[j\right]:=\left\langle\mathfrak{I}\left[j_{\mathrm{e}}=j\right]\right\rangle_{\mathbb{P}_{F}\left[\chi_{T}\right]}$, has been shown 
to satisfy the large deviation principle $\mathbb{P}_{F}\left[j\right]\approx e^{-TI_{F}\left[j\right]}$ where its rate function has also been derived explicitly as 
\begin{equation}
I_{F}\displaystyle \left[j\right]=\sum_{x,x^{\prime}\in\Omega}\int_{0}^{\infty}d\tau^{\prime}\,j\left(x;\tau^{\prime},x^{\prime}\right)\log\frac{j\left(x;\tau^{\prime},x^{\prime}\right)}{Q\left(x;\tau^{\prime}|x^{\prime}\right)g\left(x^{\prime}\right)},\label{Ij3}
\end{equation}
where $j\left(x;\tau^{\prime},x^{\prime}\right)$ should satisfy a shift-invariant property: 
\begin{equation}
g\displaystyle \left(x^{\prime}\right)=\sum_{x\in\Omega}\int_{0}^{\infty}d\tau^{\prime}\,j\left(x;\tau^{\prime},x^{\prime}\right)=\sum_{x\in\Omega}\int_{0}^{\infty}d\tau^{\prime}\,j\left(x^{\prime};\tau^{\prime},x\right).\label{shiftgjadd}
\end{equation}
If not, $I_{F}\left[j\right]$ is defined as $ I_{F}\left[j\right]=\infty$. 
A derivation of the rate function (\ref{Ij3}) and a detailed explanation for the large deviation theory are shown in Refs. \cite{18,22}. 

Next, we derive the rate function for $\mathbb{P}_{B}\left[\chi_{T}\right]$ via Eq. (\ref{Pbpath}). 
By using $j_{\mathrm{e}}\left(x;\tau^{\prime},x^{\prime}\right)$, the summation of the growth kernel $k\left(x:\tau^{\prime},x^{\prime}\right)$ in Eq. (\ref{defk}) can be represented by an integral: 
\begin{eqnarray}
\displaystyle \nonumber&&\sum_{i=1}^{n}k\left(x_{i+1};\tau_{i},x_{i}\right)\\
&&=T\displaystyle \sum_{x,x^{\prime}\in\Omega}\int_{0}^{\infty}d\tau^{\prime}\,k\left(x:\tau^{\prime},x^{\prime}\right)j_{e}\left(x;\tau^{\prime},x^{\prime}\right).\label{kje}
\end{eqnarray}
By substituting Eq. (\ref{kje}) into Eq. (\ref{Pbpath}) and averaging both sides of Eq. (\ref{Pbpath}) with $\mathfrak{I}\left[j_{\mathrm{e}}=j\right]$, 
we can obtain a coarse-grained relation between $\mathbb{P}_{F}\left[j\right]$ and $\mathbb{P}_{B}\left[j\right]$ for a sufficiently large $T$ as 
\begin{equation}
\mathbb{P}_{B}\left[j\right]\approx e^{\Lambda_{T}}e^{T\sum_{x,x^{\prime}\in\Omega}\int_{0}^{\infty}d\tau^{\prime}\,k\left(x;\tau^{\prime},x^{\prime}\right)j\left(x;\tau^{\prime},x^{\prime}\right)}\mathbb{P}_{F}\left[j\right],\label{PrBj}
\end{equation}
where $\mathbb{P}_{B}\left[j\right]:=\left\langle\mathfrak{I}\left[j_{\mathrm{e}}=j\right]\right\rangle_{\mathbb{P}_{B}\left[\chi_{T}\right]}$. 
In Eq. (\ref{PrBj}), we ignored the initial and final parts, $-\int_{0}^{\tau_{n+1}} \gamma\left(a,x_{n+1}\right)da$ and $\int_{0}^{{a_{0}}} \gamma\left(a,x_{1}\right)da$ in $\mathbb{K}\left[\chi_{T}\right]$ of Eq. (\ref{Kpathchi}), 
because these terms do not contribute to the equation for $ T\rightarrow\infty$. 
By substituting Eq. (\ref{Ij3}) and taking the limit of $ T\rightarrow\infty$ for the logarithm of the both sides of Eq. (\ref{PrBj}), we obtain 
\begin{equation}
I_{B}\displaystyle \left[j\right]=\lambda-\sum_{x,x^{\prime}\in\Omega}\int_{0}^{\infty}d\tau^{\prime}\,k\left(x;\tau^{\prime},x^{\prime}\right)j\left(x;\tau^{\prime},x^{\prime}\right)+I_{F}\left[j\right],\label{FBF}
\end{equation}
where $I_{B}\left[j\right]$ denotes the rate function for $\mathbb{P}_{B}\left[j\right]$, that is $\mathbb{P}_{B}\left[j\right]\approx e^{-TI_{B}\left[j\right]}$, and $\lambda$ is the stationary population growth rate: 
$\lambda= \lim_{T\rightarrow\infty} \Lambda_{T}/T$ (also see Eq. (\ref{SPGR})). 

This relation Eq. (\ref{FBF}) constitutes the foundation of our study, 
which represents that the rate function for the retrospective history of the population dynamics is evaluated by biasing the rate function of the semi-Markov single-cell process composed of the inter-division distribution $\pi\left(\tau|x\right)$ and the type transition matrix $\mathbb{T}\left(x|\tau^{\prime},x^{\prime}\right)$; 
furthermore, we find that this bias is determined by the growth kernel $k\left(x;\tau^{\prime},x^{\prime}\right)$. 

\section{VI. Variational structure for the stationary population growth rate}
Let us minimize Eq. (\ref{FBF}) with respect to $j\left(x;\tau^{\prime},x^{\prime}\right)$; we then find that the stationary population growth rate is given by a variational principle: 
\begin{equation}
\displaystyle \lambda=\max_{j}\left\{\sum_{x,x^{\prime}\in\Omega}\int_{0}^{\infty}d\tau^{\prime}\,k\left(x;\tau^{\prime},x^{\prime}\right)j\left(x;\tau^{\prime},x^{\prime}\right)-I_{F}\left[j\right]\right\},\label{VP}
\end{equation}
where we use the property of the rate function: $\min_{j} I_{B}\left[j\right]=0$. 
This variational form represents that the stationary population growth rate is evaluated by the Legendre transformation of the rate function for the single-cell dynamics. 
Furthermore, a maximizer of the variational form, Eq. (\ref{VP}), has an important meaning as follows. 
The maximizer of Eq. (\ref{VP}), $j_{B}^{*}\left(x;\tau^{\prime},x^{\prime}\right)$, satisfies $I_{B}\left[j_{B}^{*}\right]=0$, and therefore it represents the typical triplet on a retrospective history. 
Thereby, $j_{B}^{*}\left(x;\tau^{\prime},x^{\prime}\right)$ can be observed by a long term tracking experiment such as one by the dynamics cytometer. 
To be more precise, even if we arbitrary choose a cell in the final population and trace its ancestors back, 
we can obtain the unique triplet $j_{B}^{*}\left(x;\tau^{\prime},x^{\prime}\right)$ if the history is sufficiently long. 
The explicit form of $j_{B}^{*}\left(x;\tau^{\prime},x^{\prime}\right)$ is calculated in Appendix D. 
By using this triplet $j_{B}^{*}\left(x;\tau^{\prime},x^{\prime}\right)$, we can evaluate the stationary population growth rate as 
\begin{eqnarray}
\displaystyle \nonumber&&\lambda=\sum_{x,x^{\prime}\in\Omega}\int_{0}^{\infty}d\tau^{\prime}\,k\left(x;\tau^{\prime},x^{\prime}\right)j_{B}^{*}\left(x;\tau^{\prime},x^{\prime}\right)\\
&&-\displaystyle \sum_{x,x^{\prime}}\int_{0}^{\infty}d\tau^{\prime}\,j_{B}^{*}\left(x;\tau^{\prime},x^{\prime}\right)\log\frac{j_{B}^{*}\left(x;\tau^{\prime},x^{\prime}\right)}{Q\left(x;\tau^{\prime}|x^{\prime}\right)g_{B}^{*}\left(x^{\prime}\right)},\label{SGRj}
\end{eqnarray}
where we use the explicit form of the rate function, Eq. (\ref{Ij3}), and $g_{B}^{*}\left(x^{\prime}\right)$ is defined by Eq. (\ref{shiftgjadd}). 

\section{VII. Response relation}
By using Eq. (\ref{SGRj}), we can obtain the response of the stationary population growth rate.
First, we consider a variation of $\lambda$ with respect to the growth kernel $k\left(x;\tau^{\prime},x^{\prime}\right)$ and the semi-Markov kernel $Q\left(x;\tau^{\prime}|x^{\prime}\right)$. 
Taking Eq. (\ref{SGRj}) into account, we obtain
\begin{eqnarray}
\displaystyle \nonumber\delta\lambda&=&\displaystyle \sum_{x,x^{\prime}\in\Omega}\int_{0}^{\infty}d\tau^{\prime}\,j_{B}^{*}\left(x;\tau^{\prime},x^{\prime}\right)\delta k\left(x;\tau^{\prime},x^{\prime}\right)\\
\displaystyle \nonumber&&+\sum_{x,x^{\prime}\in\Omega}\int_{0}^{\infty}d\tau^{\prime}\,j_{B}^{*}\left(x;\tau^{\prime},x^{\prime}\right)\delta\log Q\left(x;\tau^{\prime}|x^{\prime}\right).\\
\label{dlkQ}
\end{eqnarray}
Here, the implicit variation of $\lambda$ through $j_{B}^{*}\left(x;\tau^{\prime},x^{\prime}\right)$ vanishes, because $j_{B}^{*}\left(x;\tau^{\prime},x^{\prime}\right)$ is the maximizer of Eq. (\ref{VP}), that is $\delta\lambda/\delta j_{B}^{*}\left(x;\tau^{\prime},x^{\prime}\right)=0$. 
Next, we calculate $\delta k\left(x;\tau^{\prime},x^{\prime}\right)$ and $\delta Q\left(x;\tau^{\prime}|x^{\prime}\right)$ as follows. 
By using the definition of $k\left(x;\tau^{\prime},x^{\prime}\right)$, Eq. (\ref{defk}), we have
\begin{eqnarray}
\nonumber\delta k\left(x;\tau^{\prime},x^{\prime}\right)&=&\delta\log q\left(x;\tau^{\prime},x^{\prime}\right)+\delta\log b\left(\tau^{\prime},x^{\prime}\right)\\
&&-\displaystyle \int_{0}^{\tau^{\prime}}\delta\gamma\left(t,x^{\prime}\right)dt.\label{dkbg}
\end{eqnarray}
On the other hand, from the definition of $Q\left(x;\tau^{\prime}|x^{\prime}\right)$, Eq. (\ref{defQ}), we get
\begin{equation}
\delta\log Q\left(x;\tau^{\prime}|x^{\prime}\right)=\delta\log\pi\left(\tau^{\prime}|x^{\prime}\right)+\delta\log \mathbb{T}\left(x|\tau^{\prime},x^{\prime}\right).\label{dQrT}
\end{equation}
The perturbations $\delta\log\pi\left(\tau^{\prime}|x^{\prime}\right)$ and $\delta\log \mathbb{T}\left(x|\tau^{\prime},x^{\prime}\right)$ are restricted by the stochastic conditions of $\pi\left(\tau^{\prime}|x^{\prime}\right)$ and $\mathbb{T}\left(x|\tau^{\prime},x^{\prime}\right)$: 
$\int_{0}^{\infty} \pi\left(\tau^{\prime}|x^{\prime}\right)d\tau^{\prime}=1$ and $\Sigma_{x}\mathbb{T}\left(x|\tau^{\prime},x^{\prime}\right)=1$, respectively. 
Substituting Eqs. (\ref{dkbg}) and (\ref{dQrT}) into Eq. (\ref{dlkQ}), we obtain the response relation: 
\begin{widetext}
\begin{eqnarray}
\displaystyle \nonumber\delta\lambda&=&\displaystyle \sum_{x,x^{\prime}\in\Omega}\int_{0}^{\infty}dt\,j_{B}^{*}\left(x;t,x^{\prime}\right)\delta\log q\left(x;t,x^{\prime}\right)+\sum_{x\in\Omega}\int_{0}^{\infty}dt\,g_{B}^{*}\left(t,x\right)\delta\log b\left(t,x\right)-\sum_{x\in\Omega}\int_{0}^{\infty}dt\,\mu_{B}^{*}\left(t,x\right)\delta\gamma\left(t,x\right)\\
&&+\displaystyle \sum_{x\in\Omega}\int_{0}^{\infty}dt\,g_{B}^{*}\left(t,x\right)\delta\log\pi\left(t|x\right)+\sum_{x,x^{\prime}\in\Omega}\int_{0}^{\infty}dt\,j_{B}^{*}\left(x;t,x^{\prime}\right)\delta\log \mathbb{T}\left(x|t,x^{\prime}\right),\label{RR}
\end{eqnarray}
\end{widetext}
where, to derive the third term in Eq. (\ref{RR}), we use the following property of integration: 
\begin{equation}
\displaystyle \int_{0}^{\infty}d\tau\,f\left(\tau\right)\int_{0}^{\tau}dt\,g\left(t\right)=\int_{0}^{\infty}dt\,g\left(t\right)\int_{t}^{\infty}d\tau\,f\left(\tau\right).
\end{equation}
Also, $g_{B}^{*}\left(\tau,x\right)$ and $\mu_{B}^{*}\left(a,x\right)$ represent marginal distributions of $j_{B}^{*}\left(x;\tau^{\prime},x^{\prime}\right)$: 
\begin{eqnarray}
g_{B}^{*}\displaystyle \left(\tau^{\prime},x^{\prime}\right)&:=&\displaystyle \sum_{x\in\Omega}j_{B}^{*}\left(x;\tau^{\prime},x^{\prime}\right),\\
\displaystyle \mu_{B}^{*}\left(a,x\right)&:=&\displaystyle \int_{a}^{\infty}d\tau\,g_{B}^{*}\left(\tau,x\right).
\end{eqnarray}
The marginal distribution $g_{B}^{*}\left(\tau,x\right)$ counts how often inter-division interval $\tau$ with type $x$ appears in a sufficient long retrospective history. 
On the other hand, $\mu_{B}^{*}\left(a,x\right)$ expresses the occupation density of a set $\left(a,x\right)$ on the history (see Ref. \cite{18}). 
Equation (\ref{RR}) represents the responses for any parameter change (e.g. the expected number of the daughter cells $b\left(\tau,x\right)$, the death rate $\gamma\left(a,x\right)$, and so on). 
The response coefficients in the first line of Eq .(\ref{RR}) represent responses with respect to the changes in the growth kernel. 
In contrast, those in the second line are the response coefficients for the single-cell dynamics. 
All response coefficients can be evaluated only by the typical statistics on the retrospective history. 
Therefore, we can estimate them by a long-term culturing experiment, e.g., by the dynamics cytometer \cite{11,12}. 

\section{VIII. Retrospective process}
Finally, we derive the explicit form of the time-backward rate function $I_{B}\left[j\right]$ and show that the retrospective history can be generated by another semi-Markov process, the kernel of which is characterized by $Q_{B}\left(x;\tau^{\prime}|x^{\prime}\right)$. 
We call this process the retrospective process. 
From Eq. (\ref{FBF}), we obtain 
\begin{eqnarray}
\displaystyle \nonumber I_{B}\left[j\right]&=&\displaystyle \sum_{x,x^{\prime}\in\Omega}\int_{0}^{\infty}d\tau^{\prime}\,j\left(x;\tau^{\prime},x^{\prime}\right)\\
&&\displaystyle \times\log\frac{j\left(x;\tau^{\prime},x^{\prime}\right)}{e^{k\left(x;\tau^{\prime},x^{\prime}\right)-\lambda\tau^{\prime}}Q\left(x;\tau^{\prime}|x^{\prime}\right)g\left(x^{\prime}\right)},\label{IjB3}
\end{eqnarray}
where we use the normalization condition $\Sigma_{x,x^{\prime}\in\Omega} \int_{0}^{\infty} d\tau^{\prime}\,\tau^{\prime}j\left(x;\tau^{\prime},x^{\prime}\right)=1$. 

Next, we define $u_{0}\left(0,x\right)$ by the left eigenvector corresponding to the unit eigenvalue of the matrix: 
\begin{equation}
M_{\lambda}\displaystyle \left(x|x^{\prime}\right):=\int_{0}^{\infty}d\tau^{\prime}\,e^{k\left(x;\tau^{\prime},x^{\prime}\right)-\lambda\tau^{\prime}}Q\left(x;\tau^{\prime}|x^{\prime}\right).\label{Mtext}
\end{equation}
This matrix naturally appears in a derivation of the stationary solution of the McKendric equation (\ref{McPD}) as follows. 
Assume the stationary growing condition, $N_{t}\left(a,x\right)=e^{\lambda t}v_{0}\left(a,x\right)$; then, the McKendric equation (\ref{McPD}) can be represented as 
\begin{equation}
\lambda v_{0}\left(a,x\right)=\left[-\frac{\partial}{\partial a}-\left\{\gamma\left(a,x\right)+r\left(a,x\right)\right\}\right]v_{0}\left(a,x\right).\label{stMC}
\end{equation}
From the boundary condition, Eq. (\ref{bMcPD}), $v_{0}\left(a,x\right)$ should satisfy 
\begin{eqnarray}
\displaystyle \nonumber v_{0}\left(0,x\right)&=&\displaystyle \sum_{x^{\prime}\in\Omega}\int_{0}^{\infty}d\tau^{\prime}\,q\left(x;\tau^{\prime},x^{\prime}\right)\mathbb{T}\left(x|\tau^{\prime},x^{\prime}\right)\\
&&\times b\left(\tau^{\prime},x^{\prime}\right)r\left(\tau^{\prime},x^{\prime}\right)v_{0}\left(a,x^{\prime}\right).\label{bstMC}
\end{eqnarray}
The general solution of Eq. (\ref{stMC}) is 
\begin{equation}
v_{0}\left(a,x\right)=v_{0}\left(0,x\right)e^{-\lambda a}e^{-\int_{0}^{a}\left\{\gamma\left(t,x\right)+r\left(t,x\right)\right\}dt}.\label{solstMC}
\end{equation}
By substituting Eq. (\ref{solstMC}) into Eq. (\ref{bstMC}), we have 
\begin{equation}
v_{0}\displaystyle \left(0,x\right)=\sum_{x^{\prime}\in\Omega}\int_{0}^{\infty}d\tau^{\prime}\,e^{k\left(x;\tau^{\prime},x^{\prime}\right)-\lambda\tau^{\prime}}Q\left(x;\tau^{\prime}|x^{\prime}\right)v_{0}\left(0,x^{\prime}\right),
\end{equation}
where we use the definitions of the semi-Markov kernel, Eq. (\ref{defQ}), and the growth kernel, Eq. (\ref{defk}). 
By using $M_{\lambda}\left(x|x^{\prime}\right)$, we get
\begin{equation}
v_{0}\displaystyle \left(0,x\right)=\sum_{x^{\prime}\in\Omega}M_{\lambda}\left(x|x^{\prime}\right)v_{0}\left(0,x^{\prime}\right).
\end{equation}
This equation indicates that the stationary fraction of the population with age $0$ and type $x,\ v_{0}\left(0,x\right)$, is given 
by the right eigenvector corresponding to the unit eigenvalue of $M_{\lambda}\left(x|x^{\prime}\right)$ up to a normalizing constant. 
In this section, we employ the left eigenvector $u_{0}\left(0,x\right)$ corresponding to $v_{0}\left(0,x\right)$, that is, $u_{0}\left(0,x\right)$ satisfies 
\begin{equation}
u_{0}\displaystyle \left(0,x^{\prime}\right)=\sum_{x\in\Omega}u_{0}\left(0,x\right)M_{\lambda}\left(x|x^{\prime}\right).
\end{equation}
The more detailed explanation for the matrix $M_{\lambda}\left(x|x^{\prime}\right)$ is shown in Appendices B and D. 

By using $u_{0}\left(0,x\right)$ and the following equality:
\begin{equation}
\displaystyle \sum_{x,x^{\prime}\in\Omega}\int_{0}^{\infty}d\tau^{\prime}\,j\left(x;\tau^{\prime},x^{\prime}\right)\log\frac{u_{0}\left(0,x^{\prime}\right)}{u_{0}\left(0,x\right)}=0,
\end{equation}
which is derived by the shift-invariant property of $j\left(x;\tau^{\prime},x^{\prime}\right)$, we can rewrite $I_{B}\left[j\right]$ as
\begin{equation}
I_{B}\displaystyle \left[j\right]=\sum_{x,x^{\prime}\in\Omega}\int_{0}^{\infty}d\tau^{\prime}\,j\left(x;\tau^{\prime},x^{\prime}\right)\log\frac{j\left(x;\tau^{\prime},x^{\prime}\right)}{Q_{B}\left(x;\tau^{\prime}|x^{\prime}\right)g\left(x^{\prime}\right)},\label{IBjex}
\end{equation}
where $Q_{B}\left(x;\tau^{\prime}|x^{\prime}\right)$ is defined as 
\begin{eqnarray}
\nonumber&& Q_{B}\left(x;\tau^{\prime}|x^{\prime}\right):=\\
&&u_{0}\displaystyle \left(0,x\right)e^{k\left(x;\tau^{\prime},x^{\prime}\right)-\lambda\tau^{\prime}}Q\left(x;\tau^{\prime}|x^{\prime}\right)\frac{1}{u_{0}\left(0,x^{\prime}\right)}.\label{kerBW}
\end{eqnarray}
Here, we note that $Q_{B}\left(x;\tau^{\prime}|x^{\prime}\right)$ satisfies the property of the semi-Markov kernel: 
\begin{equation}
\displaystyle \sum_{x\in\Omega}\int_{0}^{\infty}d\tau^{\prime}\,Q_{B}\left(x;\tau^{\prime}|x^{\prime}\right)=1,
\end{equation}
for any $x^{\prime}$, which is proved by the fact that $u_{0}\left(0,x\right)$ is the left eigenvector of $M_{\lambda}\left(x|x^{\prime}\right)$ with unit eigenvalue. 
This result represents that the large deviation property for the retrospective history can be mimicked 
by that for the semi-Markov process with the kernel $Q_{B}\left(x;\tau^{\prime}|x^{\prime}\right)$ in Eq. (\ref{kerBW}), 
which is an extension of the retrospective process introduced in Refs. \cite{16,17} to the MTASP. 

Since the retrospective process includes the growing effect of the population, we can reduce the calculation for a statistics on a cell-lineage tree to that on a realization of the retrospective semi-Markov process. 
This fact plays an important role in a statistical inference based on the cell-lineage tree \cite{a5}.  

Before closing this section, we further derive a reduced expression of the retrospective process by considering the case that the type transition matrix $\mathbb{T}$ and the mortality risk $q$ upon the transition are not dependent on age $\tau^{\prime}$: $\mathbb{T}\left(x|\tau^{\prime},x^{\prime}\right)=\mathbb{T}\left(x|x^{\prime}\right)$ and $q\left(x;\tau^{\prime},x^{\prime}\right)=q\left(x;x^{\prime}\right)$; 
this property is known as the direction time independence (DTI) \cite{18,19,20}. 
For the DTI case, we can get a more useful expression of the retrospective kernel than Eq. (\ref{kerBW}). 
From Eq. (\ref{defQ}), we can calculate the probability density function of the inter-division interval for the retrospective process as 
\begin{eqnarray}
\displaystyle \nonumber\pi_{B}\left(\tau^{\prime}|x^{\prime}\right)&=&\displaystyle \sum_{x\in\Omega}Q_{B}\left(x;\tau^{\prime}|x^{\prime}\right)\\
&=&\displaystyle \frac{b\left(\tau^{\prime},x^{\prime}\right)e^{-\int_{0}^{\tau^{\prime}}\gamma\left(a,x^{\prime}\right)da-\lambda\tau^{\prime}}\pi\left(\tau^{\prime}|x^{\prime}\right)}{Z\left(x^{\prime}\right)},\label{piBadd}
\end{eqnarray}
where, by using Eq. (\ref{kerBW}) and the definition of the growth kernel, Eq. (\ref{defk}), $Z\left(x^{\prime}\right)$ is evaluated as 
\begin{equation}
Z\displaystyle \left(x^{\prime}\right)=\frac{u_{0}\left(0,x^{\prime}\right)}{\sum_{x\in\Omega}u_{0}\left(0,x\right)q\left(x;x^{\prime}\right)\mathbb{T}\left(x|x^{\prime}\right)}.\label{Zu}
\end{equation}
By focusing on the form of Eq. (\ref{piBadd}), we can regard $Z\left(x^{\prime}\right)$ as the normalizing constant with respect to $\tau^{\prime}$; 
thus, we get another representation of $Z\left(x^{\prime}\right)$ as 
\begin{equation}
Z\displaystyle \left(x^{\prime}\right)=\int_{0}^{\infty}d\tau^{\prime}\,b\left(\tau^{\prime},x^{\prime}\right)e^{-\int_{0}^{\tau^{\prime}}\gamma\left(a,x^{\prime}\right)da-\lambda\tau^{\prime}}\pi\left(\tau^{\prime}|x^{\prime}\right).\label{Zpi}
\end{equation}
(Note that, if we do not assume the DTI, we can not regard $Z$ as the normalizing constant, because $Z$ is dependent on age $\tau^{\prime}$.) 
This result suggests that, for the DTI case, $\pi\left(\tau^{\prime}|x^{\prime}\right)$ can be transformed into $\pi_{B}\left(\tau^{\prime}|x^{\prime}\right)$ without knowing the type transition matrix $\mathbb{T}\left(x|x^{\prime}\right)$, if the stationary population growth rate $\lambda$ is given. 
Since $\lambda$ can be easily measured in experiments, this formula contributes to the analysis for experimental data \cite{a5}. 
On the other hand, the type transition matrix $\mathbb{T}$ is transformed as 
\begin{eqnarray}
\displaystyle \nonumber\mathbb{T}_{B}\left(x|x^{\prime}\right)&=&\displaystyle \frac{Q_{B}\left(x;\tau^{\prime}|x^{\prime}\right)}{\pi_{B}\left(\tau^{\prime}|x^{\prime}\right)}\\
\displaystyle \nonumber&=&u_{0}\displaystyle \left(0,x\right)q\left(x;x^{\prime}\right)\mathbb{T}\left(x|x^{\prime}\right)Z\left(x^{\prime}\right)\frac{1}{u_{0}\left(0,x^{\prime}\right)}\\
&=&\displaystyle \frac{u_{0}\left(0,x\right)q\left(x;x^{\prime}\right)\mathbb{T}\left(x|x^{\prime}\right)}{\sum_{x\in\Omega}u_{0}\left(0,x\right)q\left(x;x^{\prime}\right)\mathbb{T}\left(x|x^{\prime}\right)},\label{TBadd}
\end{eqnarray}
where we use Eq. (\ref{Zu}). 

\section{IX. Numerical simulation}
In order to demonstrate the analytical results derived in the previous sections, we conducted a numerical simulation. 
In this section, we assume that a cell divides into two daughters: $b\left(\tau,x\right)=2$ for any $\tau$ and $x$, and the type of the cells is in either red or blue state, $x\in\left\{\mathcal{R},\mathcal{B}\right\}$; 
we ignore the mortality risk upon the transition: $q\left(x;\tau^{\prime},x^{\prime}\right)=1$ for simplicity. 
Both the inter-division interval $\pi\left(\tau|x\right)$ and the death time distributions are modeled by the gamma distributions as 
\begin{eqnarray}
\displaystyle \pi\left(\tau|x\right)&=&\displaystyle \frac{\beta\left(x\right)^{-\alpha\left(x\right)}\tau^{\alpha\left(x\right)-1}}{\Gamma\left(\alpha\left(x\right)\right)}e^{-\tau/\beta\left(x\right)},\label{Gp}\\
\displaystyle \pi_{\gamma}\left(\tau|x\right)&=&\displaystyle \frac{\beta_{\gamma}\left(x\right)^{-\alpha_{\gamma}\left(x\right)}\tau^{\alpha_{\gamma}\left(x\right)-1}}{\Gamma\left(\alpha_{\gamma}\left(x\right)\right)}e^{-\tau/\beta_{\gamma}\left(x\right)},
\end{eqnarray}
where $\Gamma\left(\alpha\right)$ is Euler's gamma function and $\pi_{\gamma}\left(\tau|x\right)$ is related to the death rate $\gamma\left(a,x\right)$ as $\pi_{\gamma}\left(\tau|x\right)=\gamma\left(\tau,x\right)e^{-\int_{0}^{\tau}\gamma\left(a,x\right)da}$; that is 
\begin{equation}
\displaystyle \gamma\left(a,x\right)=\frac{\pi_{\gamma}\left(a|\bm{x}\right)}{\int_{a}^{\infty}\pi_{\gamma}\left(\tau|\bm{x}\right)d\tau}.\label{G1p}
\end{equation}
On average, the red type is assumed to have shorter inter-division interval and death time than those of the blue type, and the parameters of the corresponding gamma distributions are specified as 
\begin{eqnarray}
\nonumber\left\{\alpha\left(\mathcal{R}\right),\beta\left(\mathcal{R}\right)\right\}&=&\left\{4,0.5\right\},\,\,\,\,\left\{\alpha\left(\mathcal{B}\right),\beta\left(\mathcal{B}\right)\right\}=\left\{4,1\right\},\\
\nonumber\left\{\alpha_{\gamma}\left(\mathcal{R}\right),\beta_{\gamma}\left(\mathcal{R}\right)\right\}&=&\left\{8,0.5\right\},\,\,\,\,\left\{\alpha_{\gamma}\left(\mathcal{B}\right),\beta_{\gamma}\left(\mathcal{B}\right)\right\}=\left\{8,1\right\}.\\\label{setup1}
\end{eqnarray}
The stochastic transition matrix of the types is age-independently set to be 
\begin{equation}
\mathbb{T}=\left(\begin{array}{l}
0.8\,\,\,\,0.2\\
0.2\,\,\,\,0.8
\end{array}\right).\label{setup2}
\end{equation}
Figure 3 (A) illustrates the outline of the above model. 
\begin{figure}[h]
\begin{center}
\includegraphics*[height=10.5cm]{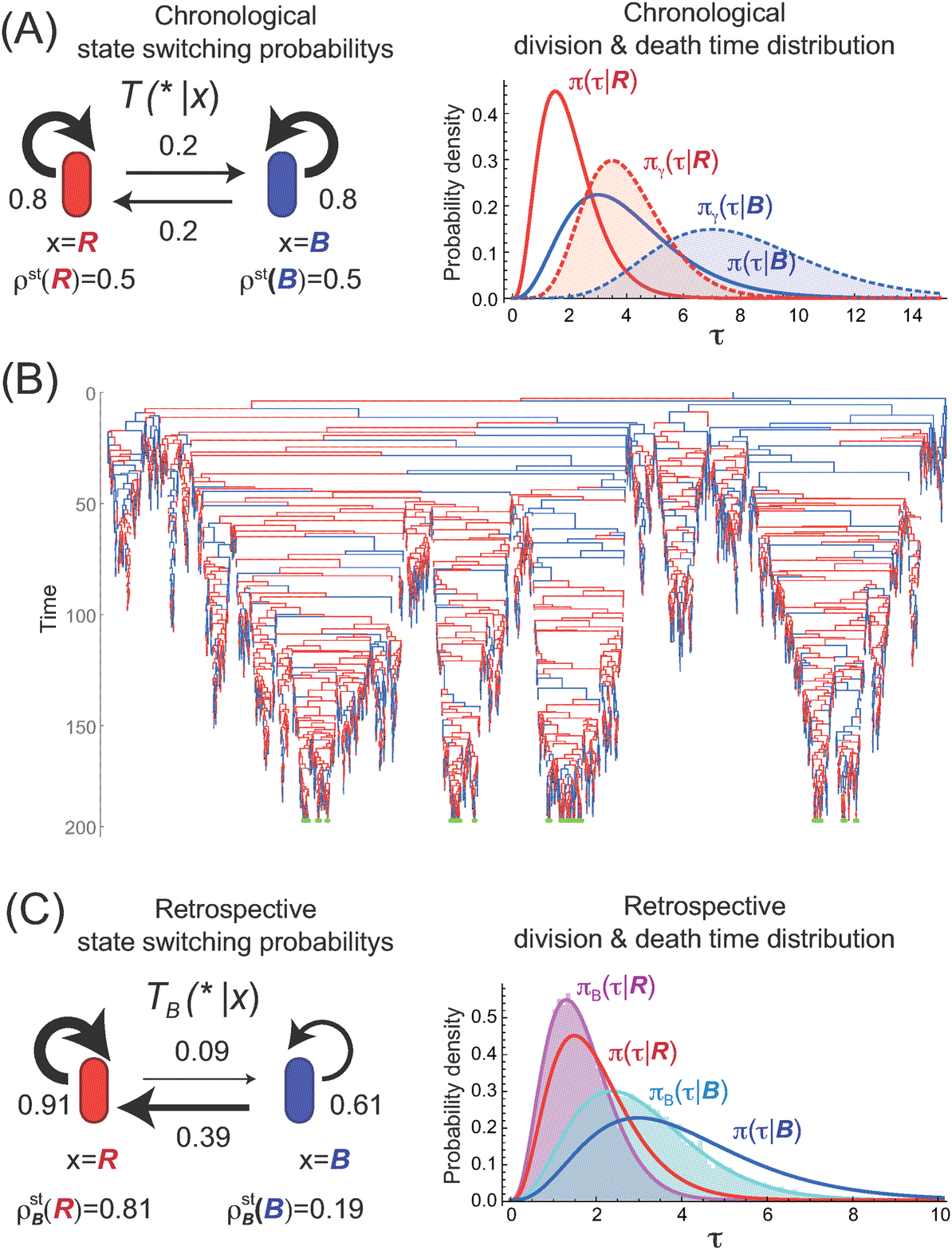}
\caption{(A) A schematic diagram of the two state model of the cells used in the simulation. 
The left panel shows the values of the state-switching probability and its stationary distribution: $\Sigma_{x^{\prime}\in\Omega}\mathbb{T}\left(x|x^{\prime}\right)\rho^{\mathrm{s}\mathrm{t}}\left(x^{\prime}\right)=\rho^{\mathrm{s}\mathrm{t}}\left(x\right)$. 
The right panel represents the distributions of the inter-division intervals and the death intervals for the two states of the cells. 
(B) A lineage tree obtained by the conducted simulation. 
The first $10^{4}$ cells are shown in the lineage. 
(C) A schematic diagram of the state-switching probability, the distribution of the inter-division interval, and that of the death interval for the retrospective process. 
The left panel shows the values of the transition probability for the retrospective process and the corresponding stationary distribution: $\Sigma_{x^{\prime}\in\Omega}\mathbb{T}_{B}\left(x|x^{\prime}\right)\rho_{B}^{\mathrm{s}\mathrm{t}}\left(x^{\prime}\right)=\rho_{B}^{\mathrm{s}\mathrm{t}}\left(x\right)$. 
In the right panel, the solid curves are calculated by the analytic expressions, Eq. (\ref{gmpiB}), and the histograms are empirically obtained from a retrospective path from the lineage tree.}
\end{center}
\end{figure}

To approximately simulate the branching process defined in Sec. III, we consider the population of the cells cultured in the dynamics cytometer (see FIG. 2 (A)) that has a limited chamber size. 
We set the capacity of the camber to be $N_{\mathrm{m}\mathrm{a}\mathrm{x}}=100$. 
In the dynamics cytometer, the cells exceeding the capacity is washed out by the flowing medium at the boundaries of the chamber. 
To mimic this property, when the population size exceeds the capacity by the division event of a cell in our simulation, we randomly choose one cell in the chamber including the newborn cells, 
and remove it from the chamber so that the population size is kept to be no more than the capacity. 
This assumption is almost equivalent to introducing an age- and type-independent constant death rate $\gamma_{\mathrm{f}\mathrm{l}\mathrm{o}\mathrm{w}}$ that balances the stationary population growth rate of the cells as $\lambda=\gamma_{\mathrm{f}\mathrm{l}\mathrm{o}\mathrm{w}}$. 
Owing to this property, we can estimate the stationary population growth rate $\lambda$ by counting the number of the flown cells from the chamber up to time $t,\ N_{\mathrm{f}\mathrm{l}\mathrm{o}\mathrm{w}}\left(t\right)$, as 
\begin{equation}
\displaystyle \lambda=\lim_{t\rightarrow\infty}\frac{1}{N_{\mathrm{m}\mathrm{a}\mathrm{x}}}\frac{N_{\mathrm{f}\mathrm{l}\mathrm{o}\mathrm{w}}\left(t\right)}{t}.\label{lamflo}
\end{equation}

Figure 3 (B) is a cell-lineage tree obtained by conducting the simulation up to the time when we have $10^{4}$ new cells in the population starting from one root cell. 
In order to obtain a sufficiently long retrospective history (path) and its empirical distributions, we conducted the same simulation up to the time when $5\times 10^{6}$ new cells are generated. 
Figure 3 (C) shows the histograms of the retrospective empirical distribution for the inter-division interval calculated from a retrospective path sampled from the lineage tree: 
\begin{equation}
\displaystyle \pi_{B}\left(\tau|x\right)=\frac{\#\,\mathrm{o}\mathrm{f}\,\mathrm{c}\mathrm{e}\mathrm{l}\mathrm{l}\mathrm{s}\,\mathrm{w}\mathrm{i}\mathrm{t}\mathrm{h}\,\tau\,\mathrm{a}\mathrm{n}\mathrm{d}\,x\,\mathrm{o}\mathrm{n}\,\mathrm{t}\mathrm{h}\mathrm{e}\,\mathrm{p}\mathrm{a}\mathrm{t}\mathrm{h}}{\#\,\mathrm{o}\mathrm{f}\,\mathrm{c}\mathrm{e}\mathrm{l}\mathrm{l}\mathrm{s}\,\mathrm{w}\mathrm{i}\mathrm{t}\mathrm{h}\,\mathrm{t}\mathrm{y}\mathrm{p}\mathrm{e}\,x\,\mathrm{o}\mathrm{n}\,\mathrm{t}\mathrm{h}\mathrm{e}\,\mathrm{p}\mathrm{a}\mathrm{t}\mathrm{h}}.
\end{equation}
By following the theory developed in Sec. VIII, the empirical histograms should coincide with the following distribution obtained by using Eqs. (\ref{piBadd}), (\ref{Gp}) and (\ref{G1p}): 
\begin{eqnarray}
\displaystyle \nonumber\pi_{B}\left(\tau|x\right)&=&\displaystyle \frac{2e^{-\int_{0}^{\tau}\gamma\left(a,x\right)da-\lambda\tau}}{Z\left(x\right)}\pi\left(\tau|x\right)\\
\displaystyle \nonumber&=&\displaystyle \frac{2e^{-\lambda\tau}\int_{\tau}^{\infty}\pi_{\gamma}\left(t|x\right)dt}{Z\left(x\right)}\pi\left(\tau|x\right)\\
\displaystyle \nonumber&=&\displaystyle \frac{2}{Z\left(x\right)}\frac{\beta\left(x\right)^{-\alpha\left(x\right)}}{\Gamma\left(\alpha\left(x\right)\right)\Gamma\left(\alpha_{\gamma}\left(x\right)\right)}\\
\nonumber&&\times\Gamma\left(\alpha_{\gamma}\left(x\right),\tau/\beta_{\gamma}\left(x\right)\right)\tau^{\alpha\left(x\right)-1}e^{-\left(\lambda+1/\beta\left(x\right)\right)\tau},\\\label{gmpiB}
\end{eqnarray}
where $\Gamma\left(\alpha,\tau\right):= \int_{\tau}^{\infty} t^{a-1}e^{-t}dt$ is the upper incomplete gamma function and $Z\left(x\right)$ is the normalizing constant calculated by Eq. (\ref{Zpi}). 
As demonstrated in FIG. 3 (C),  Eq. (\ref{gmpiB}) is perfectly fitted to the empirical histograms, where we calculated Eq. (\ref{gmpiB}) by employing the parameter values, Eq. (\ref{setup1}), and the population growth rate $\lambda=0.255106$ estimated with Eq. (\ref{lamflo}). 
Similarly, we calculated the empirical $\mathbb{T}_{B}$ from the retrospective path as in TABLE I; 
\begin{equation}
\displaystyle \mathbb{T}_{B}\left(x|x^{\prime}\right)=\frac{\#\,\mathrm{o}\mathrm{f}\,\mathrm{t}\mathrm{r}\mathrm{a}\mathrm{n}\mathrm{s}\mathrm{i}\mathrm{t}\mathrm{i}\mathrm{o}\mathrm{n}\mathrm{s}\,\mathrm{f}\mathrm{r}\mathrm{o}\mathrm{m}\,x^{\prime}\,\mathrm{t}\mathrm{o}\,x\,\mathrm{o}\mathrm{n}\,\mathrm{t}\mathrm{h}\mathrm{e}\,\mathrm{p}\mathrm{a}\mathrm{t}\mathrm{h}}{\#\,\mathrm{o}\mathrm{f}\,\mathrm{t}\mathrm{r}\mathrm{a}\mathrm{n}\mathrm{s}\mathrm{i}\mathrm{t}\mathrm{i}\mathrm{o}\mathrm{n}\mathrm{s}\,\mathrm{f}\mathrm{r}\mathrm{o}\mathrm{m}\,\mathrm{t}\mathrm{y}\mathrm{p}\mathrm{e}\,x^{\prime}\,\mathrm{o}\mathrm{n}\,\mathrm{t}\mathrm{h}\mathrm{e}\,\mathrm{p}\mathrm{a}\mathrm{t}\mathrm{h}}.
\end{equation}
This empirical $\mathbb{T}_{B}$ almost perfectly agrees with that calculated from the analytical expression, Eq. (\ref{TBadd}). 
\begin{table}[htp]
\begin{center}
\begin{tabular}{|c|c|c|c||c|c|c|c|}
\hline
\multicolumn{4}{|c||}{ Empirical $\mathbb{T}_{B}(x|x')$}  & \multicolumn{4}{|c|}{ Analytical $\mathbb{T}_{B}(x|x')$}\\
\cline{1-8}
\multicolumn{2}{|c|}{}& \multicolumn{2}{|c||}{$x'$}  &\multicolumn{2}{|c|}{} &\multicolumn{2}{|c|}{$x'$}\\ \cline{3-4}\cline{7-8}
\multicolumn{2}{|c|}{}& $\mathcal{R}$ & $\mathcal{B}$ & \multicolumn{2}{|c|}{} & $\mathcal{R}$ & $\mathcal{B}$ \\ \hline
$x$&$\mathcal{R}$ & 0.910648 & 0.390727 &  $x$ &$\mathcal{R}$  & 0.911539 & 0.391355 \\ \cline{2-4}\cline{6-8}
&$\mathcal{B}$ & 0.0893523 & 0.609273 &  &$\mathcal{B}$  & 0.0886272 & 0.608811 \\\hline
\end{tabular}
\caption{The comparison of $\mathbb{T}_{B}(x|x')$ obtained empirically from a retrospective path with that obtained from the analytical expression.}\label{tb:TB}
\end{center}
\end{table}
To numerically evaluate this analytical expression, we firstly calculated the matrix $M_{\lambda}\left(x|x^{\prime}\right)$, Eq. (\ref{Mtext}), as 
\begin{eqnarray}
\displaystyle \nonumber M_{\lambda}\left(x|x^{\prime}\right)&=&\displaystyle \mathbb{T}\left(x|x^{\prime}\right)\int_{0}^{\infty}d\tau^{\prime}\,2e^{-\int_{0}^{\tau^{\prime}}\gamma\left(a,x\right)da-\lambda\tau^{\prime}}\pi\left(\tau^{\prime}|x^{\prime}\right)\\
&=&\mathbb{T}\left(x|x^{\prime}\right)Z\left(x^{\prime}\right).
\end{eqnarray}
Then, we numerically obtained the left eigenvector $u_{0}\left(0,x\right)$ of the matrix $M_{\lambda}\left(x|x^{\prime}\right)$ by using the actual value of $\mathbb{T}$, Eq. (\ref{setup2}). 
Finally, we confirmed the agreement of the empirical distribution of the type obtained from the retrospective path: 
\begin{equation}
\displaystyle \rho_{B}^{\mathrm{s}\mathrm{t}}\left(x\right)=\frac{\#\,\mathrm{o}\mathrm{f}\,\mathrm{c}\mathrm{e}\mathrm{l}\mathrm{l}\mathrm{s}\,\mathrm{w}\mathrm{i}\mathrm{t}\mathrm{h}\,x\,\mathrm{o}\mathrm{n}\,\mathrm{t}\mathrm{h}\mathrm{e}\,\mathrm{p}\mathrm{a}\mathrm{t}\mathrm{h}}{\#\,\mathrm{o}\mathrm{f}\,\mathrm{c}\mathrm{e}\mathrm{l}\mathrm{l}\mathrm{s}\,\mathrm{o}\mathrm{n}\,\mathrm{t}\mathrm{h}\mathrm{e}\,\mathrm{p}\mathrm{a}\mathrm{t}\mathrm{h}},
\end{equation}
with that of the analytical expression: $\rho_{B}^{\mathrm{s}\mathrm{t}}\left(x\right)=u_{0}\left(0,x\right)v_{0}\left(0,x\right)$, which is derived as follows. 
From Eq. (\ref{TBadd}), we obtain 
\begin{equation}
\displaystyle \sum_{x^{\prime}\in\Omega}\mathbb{T}\left(x|x^{\prime}\right)Z\left(x^{\prime}\right)\frac{\rho_{B}^{\mathrm{s}\mathrm{t}}\left(x^{\prime}\right)}{u_{0}\left(0,x^{\prime}\right)}=\frac{\rho_{B}^{\mathrm{s}\mathrm{t}}\left(x\right)}{u_{0}\left(0,x\right)},
\end{equation}
which represents that $\rho_{B}^{\mathrm{s}\mathrm{t}}\left(x\right)/u_{0}\left(0,x\right)$ corresponds to the right eigenvector of $M_{\lambda}\left(x|x^{\prime}\right)$ with unit eigenvalue, that is $\rho_{B}^{\mathrm{s}\mathrm{t}}\left(x\right)/u_{0}\left(0,x\right)=v_{0}\left(0,x\right)$. 
Thus, we get the above analytical expression. 
This empirical distribution is also known as the ancestral distribution in population genetics \cite{16,17}. 
Here, $u_{0}\left(0,x\right)$ and $v_{0}\left(0,x\right)$ were calculated by numerically solving the eigenvalue problem associated with the matrix $M_{\lambda}\left(x|x^{\prime}\right)$. 
Note that $\rho_{B}^{\mathrm{s}\mathrm{t}}\left(x^{\prime}\right)$ coincides with $g_{B}^{*}\left(x^{\prime}\right)=\left(1/T\right)\Sigma_{i=1}^{n}\delta_{x^{\prime},x_{i}}=\Sigma_{x\in\Omega} \int_{0}^{\infty} d\tau^{\prime}j_{B}^{*}\left(x;\tau^{\prime},x^{\prime}\right)$ up to a normalizing constant (see Eq. (\ref{Dapp4})). 
The details of the analytical expression, $u_{0}\left(0,x\right)v_{0}\left(0,x\right)$, is shown in Appendix D. 
As TABLE II demonstrates, both distributions are almost identical. 
\begin{table}[htp]
\begin{center}
\begin{tabular}{|c|c|c||c|c|c|}
\hline
\multicolumn{3}{|c||}{ Empirical $\rho_{B}^{\mathrm{s}\mathrm{t}}(x)$}  & \multicolumn{3}{|c|}{ Analytical $\rho_{B}^{\mathrm{s}\mathrm{t}}(x)$}\\
\hline
& \multicolumn{2}{|c||}{$x$}   & &\multicolumn{2}{|c|}{$x$}\\ \cline{2-3}\cline{5-6}
& $\mathcal{R}$ & $\mathcal{B}$ &  & $\mathcal{R}$ & $\mathcal{B}$ \\ \hline
$\rho_{B}^{\mathrm{s}\mathrm{t}}(x)$& 0.813904 & 0.186096 &  $\rho_{B}^{\mathrm{s}\mathrm{t}}(x)$  & 0.815353 & 0.184647 \\ \cline{2-3}\cline{5-6}
\hline
\end{tabular}
\caption{The comparison of $\rho_{B}^{\mathrm{s}\mathrm{t}}(x)$ obtained empirically from a retrospective path with that obtained from the analytical expression.}\label{tb:TB}
\end{center}
\label{tb:rho}
\end{table}

Next, we verified the fitness response relation, Eq. (\ref{RR}). 
Specifically, we calculated the population growth rate $\lambda$ numerically for perturbed values of the parameters, and compared the results with the response predicted by Eq. (\ref{RR}) (see FIGs. 4-6). 
In FIG. 4, we perturbed $\pi\left(\tau|x\right)$ by changing $\left\{\alpha\left(\mathcal{R}\right),\beta\left(\mathcal{R}\right)\right\}$ (FIG. 4 (A)) and $\left\{\alpha\left(\mathcal{B}\right),\beta\left(\mathcal{B}\right)\right\}$ (FIG. 4 (B)). 
Similarly, in FIG. 5, we perturbed $\gamma\left(a,x\right)$ by changing $\left\{\alpha_{\gamma}\left(\mathcal{R}\right),\beta_{\gamma}\left(\mathcal{R}\right)\right\}$ (FIG. 5 (A)) and $\left\{\alpha_{\gamma}\left(\mathcal{B}\right),\beta_{\gamma}\left(\mathcal{B}\right)\right\}$ (FIG. 5 (B)). 
For perturbing $\mathbb{T}$, we parameterized $\mathbb{T}$ with $\left\{\theta_{\mathcal{R}},\theta_{\mathcal{B}}\right\}$ as 
\begin{equation}
\mathbb{T}=\left(\begin{array}{l}
\,\,\,\,\,T_{\mathcal{R}}^{\theta_{\mathcal{R}}}\,\,\,\,\,\,\,\,\,\,\,1-T_{\mathcal{B}}^{\theta_{\mathcal{B}}}\\
1-T_{\mathcal{R}}^{\theta_{\mathcal{R}}}\,\,\,\,\,\,\,\,\,T_{\mathcal{B}}^{\theta_{\mathcal{B}}}
\end{array}\right),\label{ttheio}
\end{equation}
and we set $T_{\mathcal{R}}=T_{\mathcal{B}}=0.8$ so that Eq. (\ref{ttheio}) becomes identical to Eq. (\ref{setup2}) when $\left\{\theta_{\mathcal{R}},\theta_{\mathcal{B}}\right\}=\left\{1,1\right\}$. 
In FIG. 6, we perturbed $\mathbb{T}$ by changing $\left\{\theta_{\mathcal{R}},\theta_{\mathcal{B}}\right\}$. 
To estimate the stationary population growth rate for each parameter value, we generated a lineage tree with $5\times 10^{4}$ cells and used Eq. (\ref{lamflo}). 
From the same tree, a retrospective path was sampled, and the response coefficients, $g_{B}^{*}\left(\tau,x\right),\ \mu_{B}^{*}\left(a,x\right)$ and $j_{B}^{*}\left(x,x^{\prime}\right)= \int_{0}^{\infty} d\tau^{\prime}j_{B}^{*}\left(x;\tau^{\prime},x^{\prime}\right)$ were empirically calculated from the path (see right panels in FIGs. 4-6). 
For all the cases we have tested, the stationary growth rate exactly responds to the changes in the parameters as predicted by Eq. (\ref{RR}), which clearly demonstrate the validity of the relation. 
\begin{figure}[h]
\begin{center}
\includegraphics*[height=6cm]{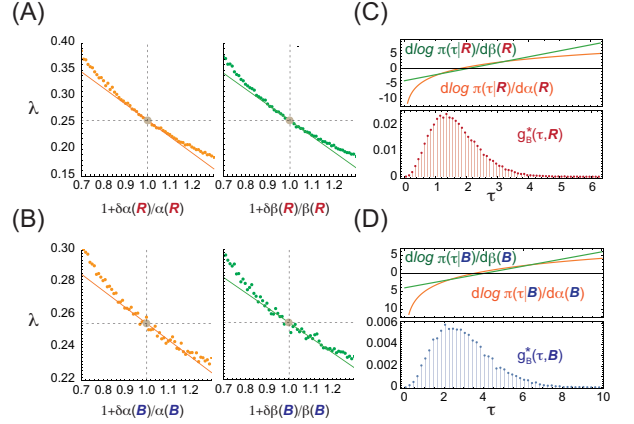}
\caption{The responses of the stationary population growth rate and the response coefficients for the perturbations to $\pi\left(\tau|\mathcal{R}\right)$ (A, C) and $\pi\left(\tau|\mathcal{B}\right)$ (B, D). 
(A, B) The actual responses of the growth rate (points) and the predicted responses (lines) respectively to the perturbations for the state $\mathcal{R}$ (A) and the state $\mathcal{B}$ (B). 
The parameters are perturbed around the same parameter values as in Eq. (\ref{setup1}). 
(C, D) The response coefficient and the changes in $\pi\left(\tau|x\right)$ induced by the parameter perturbations of $\alpha\left(\mathcal{R}\right)$ and $\beta\left(\mathcal{R}\right)$ (C) and $\alpha\left(\mathcal{B}\right)$ and $\beta\left(\mathcal{B}\right)$ (D).}
\end{center}
\end{figure}
\begin{figure}[h]
\begin{center}
\includegraphics*[height=6cm]{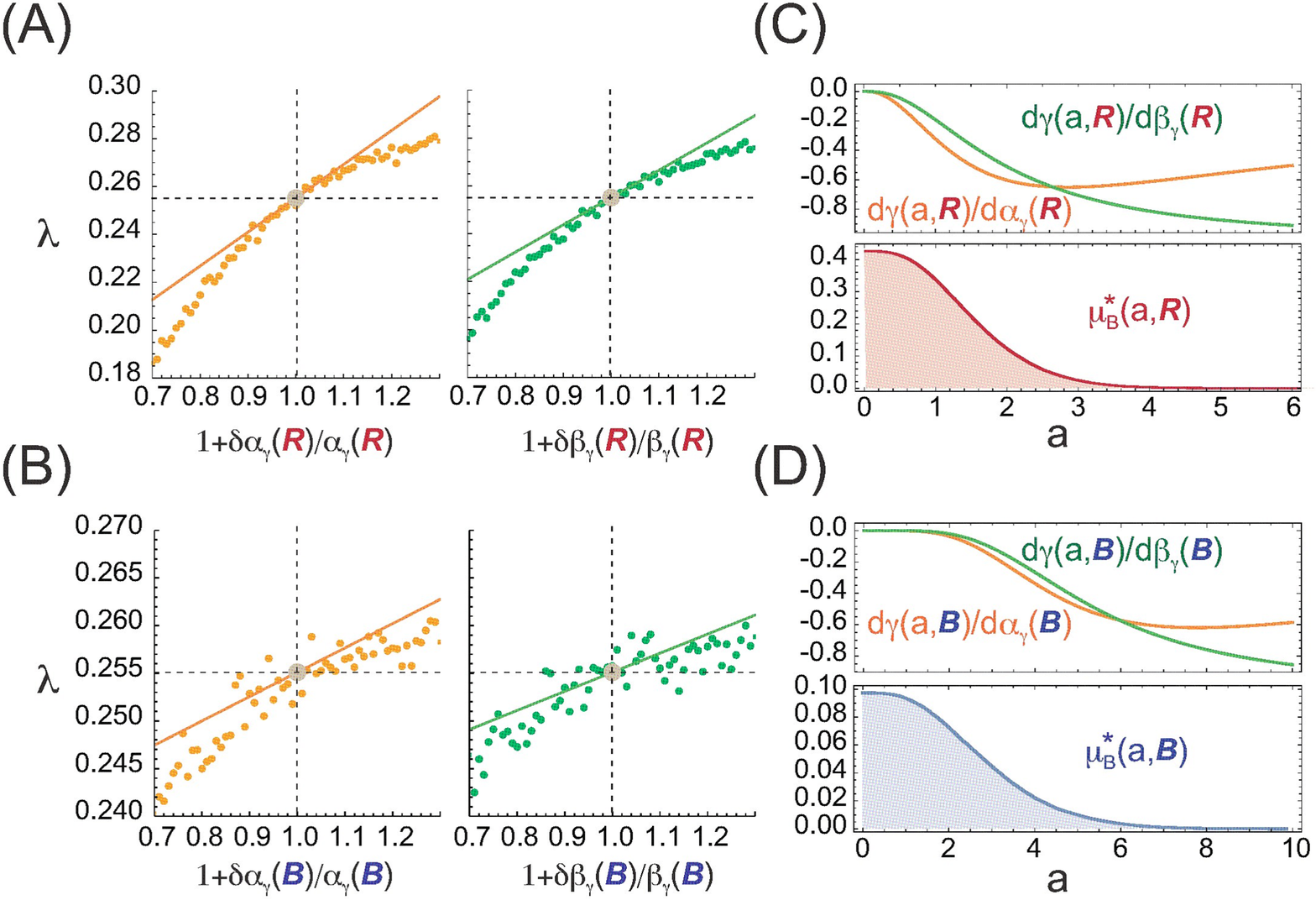}
\caption{The responses of the stationary population growth rate and the response coefficients for the perturbations to $\gamma\left(\tau,\mathcal{R}\right)$ (A, C) and $\gamma\left(\tau,\mathcal{B}\right)$ (B, D). 
(A, B) The actual responses of the growth rate (points) and the predicted responses (lines) respectively to the perturbations for the state $\mathcal{R}$ (A) and the state $\mathcal{B}$ (B). 
The parameters are perturbed around the same parameter values as in Eq. (\ref{setup1}). 
(C, D) The response coefficient and the changes in $\gamma\left(a,x\right)$ induced by the parameter perturbations of $\alpha_{\gamma}\left(\mathcal{R}\right)$ and $\beta_{\gamma}\left(\mathcal{R}\right)$ (C) and $\alpha_{\gamma}\left(\mathcal{B}\right)$ and $\beta_{\gamma}\left(\mathcal{B}\right)$ (D).}
\end{center}
\end{figure}
\begin{figure}[h]
\begin{center}
\includegraphics*[height=6cm]{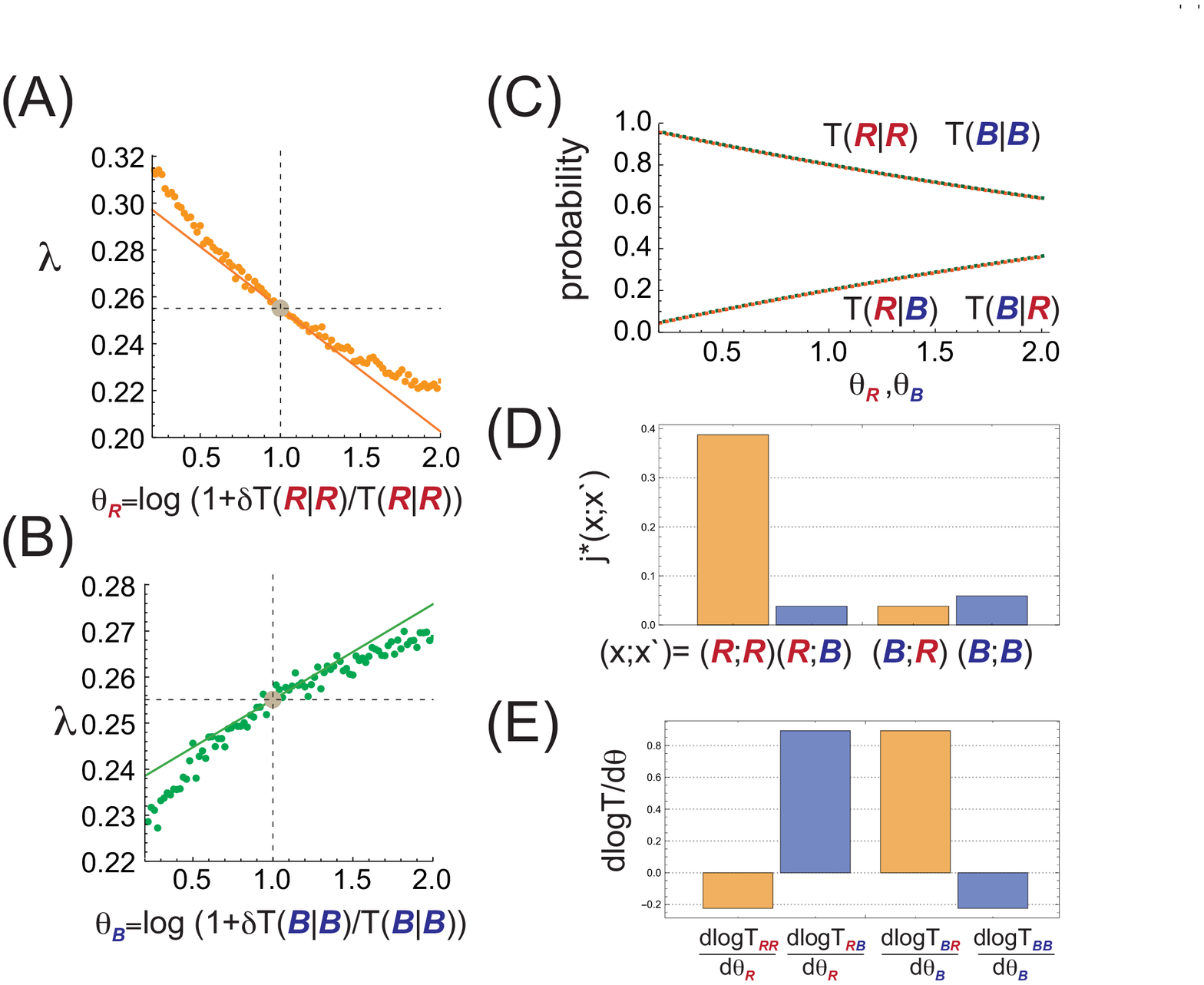}
\caption{The responses of the stationary population growth rate and the response coefficients for the perturbations to $\mathbb{T}$. 
(A, B) The actual responses of the growth rate (points) and the predicted responses (lines) respectively to the perturbations of $\theta_{\mathcal{R}}$ (A) and $\theta_{\mathcal{B}}$ (B). 
(C) The changes in the value of the components in $\mathbb{T}(x|x^{\prime})$ as a function of either $\theta_{\mathcal{R}}$ or $\theta_{\mathcal{B}}$. 
(D) The response coefficient $j_{B}^{*}\left(x,x^{\prime}\right)$. 
(E) The changes in the components of $\mathbb{T}(x|x^{\prime})$ induced by the parameter perturbations of $\theta_{\mathcal{R}}$ or $\theta_{\mathcal{B}}$. 
The components not shown in the chart are zero.}
\end{center}
\end{figure}
\section{X. Summary}
In this work, we have constructed the pathwise formulation for the MTASP. 
By employing the formulation, we have derived the variational representation of the stationary population growth rate, which comprises a trade-off between growth effects and a single-cell dynamics. 
Owing to this variational representation, a response relation of the stationary population growth rate has been obtained, in which various retrospective distributions work as the response coefficients. 
Thereby, the response can be evaluated by statistics on the retrospective history. 
The derived relations have been verified by the numerical simulations. 
Our result can be directly employed to estimate how bacteria and other cells behave in response to perturbations just by measuring the retrospective history without relying on the inference of dynamics and the eigenvalue problem. 
Moreover, our pathwise formulation and variational representation of the MTASP can also be applied for designing statistical inference algorithms of underlying parameters and dynamics from the experimentally observed lineage tree \cite{a5}. 
All these results may contribute to extending our ability to predict and control evolution.

\section{ACKNOWLEDGMENTS}
We acknowledge Yuichi Wakamoto, Edo Kussell, Takashi Nozoe and Ryo Oizumi for their useful discussions. 
This work is supported by JSPS KAKENHI Grant Numbers JP16K17763 and JP16H06155; and JST PRESTO Grant Number JPMJPR15E4. 

\appendix
\section{Appendix A}
Here, we derive the McKendric equation (\ref{McPD}) and its boundary condition (\ref{bMcPD}). 
Let $P_{t+\Delta t}^{\left(a,x\right)}\left(N\right)$ be the probability distribution of the number of the cells that have age $a$ and type $x$ at time $t+\Delta t$, which is normalized as $\Sigma_{N=0}^{\infty}P_{t+\Delta t}^{\left(a,x\right)}\left(N\right)=1$. 
First, we consider the case $a\neq 0$. 
By employing the death rate $\gamma\left(a,x\right)$ and the division rate $r\left(a,x\right)$, we can calculate $P_{t+\Delta t}^{\left(a,x\right)}\left(N\right)$ as 
\begin{eqnarray}
\nonumber&& P_{t+\Delta t}^{\left(a,x\right)}\left(N\right)=\left(N+1\right)\\
\nonumber&&\times\left\{\gamma\left(a-\Delta t,x\right)+r\left(a-\Delta t,x\right)\right\}\Delta tP_{t}^{\left(a-\Delta t,x\right)}\left(N+1\right)\\
\nonumber&&+\left[1-\left\{\gamma\left(a-\Delta t,x\right)+r\left(a-\Delta t,x\right)\right\}\Delta t\right]^{N}P_{t}^{\left(a-\Delta t,x\right)}\left(N\right).\\\label{A3}
\end{eqnarray}
Here, the first term represents the probability that the number of cells with $\left(a,x\right)$ changes from $N+1$ to $N$ due to death or division of the cells; 
on the other hand, the second term is the probability that the age of the cells shifts from $a-\Delta t$ to $a$ by aging. 
Using $\Delta t\rightarrow 0$, we can obtain the following two approximations: 
\begin{equation}
P_{t}^{\left(a-\Delta t,x\right)}\displaystyle \left(N\right)\approx P_{t}^{\left(a,x\right)}\left(N\right)-\frac{\partial P_{t}^{\left(a,x\right)}\left(N\right)}{\partial a}\Delta t,\label{A1}
\end{equation}
and
\begin{eqnarray}
\nonumber&&\left[1-\left\{\gamma\left(a-\Delta t,x\right)+r\left(a-\Delta t,x\right)\right\}\Delta t\right]^{N}\\
&\approx&1-N\left\{\gamma\left(a,x\right)+r\left(a,x\right)\right\}\Delta t.\label{A2}
\end{eqnarray}
By substituting these approximations (\ref{A1}) and (\ref{A2}) into Eq. (\ref{A3}), we obtain the stochastic time evolution equation of the number of the cells: 
\begin{eqnarray}
\displaystyle \nonumber&&\frac{\partial P_{t}^{\left(a,x\right)}\left(N\right)}{\partial t}=\left(N+1\right)\left\{\gamma\left(a,x\right)+r\left(a,x\right)\right\}P_{t}^{\left(a,x\right)}\left(N+1\right)\\
&&-\displaystyle \frac{\partial P_{t}^{\left(a,x\right)}\left(N\right)}{\partial a}-N\left\{\gamma\left(a,x\right)+r\left(a,x\right)\right\}P_{t}^{\left(a,x\right)}\left(N\right).\label{A5}
\end{eqnarray}
By using this equation and considering the time evolution of the expectation: $N_{t}\left(a,x\right):=\Sigma_{N=1}^{\infty}NP_{t}^{\left(a,x\right)}\left(N\right)$, we obtain the McKendric equation:
\begin{equation}
\displaystyle \frac{\partial}{\partial t}N_{t}\left(a,x\right)=\left[-\frac{\partial}{\partial a}-\left\{\gamma\left(a,x\right)+r\left(a,x\right)\right\}\right]N_{t}\left(a,x\right).\label{A6}
\end{equation}
Next, from the case $a=0$, we derive the boundary condition (\ref{bMcPD}). 
The probability $P_{t}^{\left(0,x\right)}\left(N\right)$ is calculated as 
\begin{eqnarray}
\displaystyle \nonumber&& P_{t}^{\left(0,x\right)}\left(N\right)=\sum_{N^{\prime}=1}^{\infty}\sum_{x^{\prime}\in\Omega}\int_{0}^{\infty}d\tau^{\prime}\,r\left(\tau^{\prime},x^{\prime}\right)N^{\prime}P_{t}^{\left(\tau^{\prime},x^{\prime}\right)}\left(N^{\prime}\right)\\
\displaystyle \nonumber&&\times\biggl[\sum_{z^{\prime}=N}^{\infty}\frac{z^{\prime}!}{N!\left(z^{\prime}-N\right)!}p\left(z^{\prime}|\tau^{\prime},x^{\prime}\right)\left(q\left(x;\tau^{\prime},x^{\prime}\right)\mathbb{T}\left(x|\tau^{\prime},x^{\prime}\right)\right)^{N}\\
\nonumber&&\times\left\{\left(1-\mathbb{T}\left(x|\tau^{\prime},x^{\prime}\right)\right)+\left(1-q\left(x;\tau^{\prime},x^{\prime}\right)\right)\mathbb{T}\left(x|\tau^{\prime},x^{\prime}\right)\right\}^{z^{\prime}-N}\biggr]\\
\displaystyle \nonumber&&=\sum_{N^{\prime}=1}^{\infty}N^{\prime}\sum_{x^{\prime}\in\Omega}\int_{0}^{\infty}d\tau^{\prime}\,r\left(\tau^{\prime},x^{\prime}\right)P_{t}^{\left(\tau^{\prime},x^{\prime}\right)}\left(N^{\prime}\right)\\
\displaystyle \nonumber&&\times\biggl[\sum_{z^{\prime}=N}^{\infty}\frac{z^{\prime}!}{N!\left(z^{\prime}-N\right)!}p\left(z^{\prime}|\tau^{\prime},x^{\prime}\right)\left(q\left(x;\tau^{\prime},x^{\prime}\right)\mathbb{T}\left(x|\tau^{\prime},x^{\prime}\right)\right)^{N}\\
&&\times\left(1-q\left(x;\tau^{\prime},x^{\prime}\right)\mathbb{T}\left(x|\tau^{\prime},x^{\prime}\right)\right)^{z^{\prime}-N}\biggr],\label{A4}
\end{eqnarray}
where the inside of the bracket $\left[\cdot\right]$ consists of the product of two probabilities: 
The first one represents the probability that $N$ cells in $z^{\prime}$ daughters succeed in the type transition from type $x^{\prime}$ to $x$, 
that is $\left(q\left(x;\tau^{\prime},x^{\prime}\right)\mathbb{T}\left(x|\tau^{\prime},x^{\prime}\right)\right)^{N}$. 
In contrast, the second one is the probability that $z^{\prime}-N$ cells in $z^{\prime}$ daughters switch to a different type from $x^{\prime}$ or fail the type transition: 
$\left\{\left(1-\mathbb{T}\left(x|\tau^{\prime},x^{\prime}\right)\right)+\left(1-q\left(x;\tau^{\prime},x^{\prime}\right)\right)\mathbb{T}\left(x|\tau^{\prime},x^{\prime}\right)\right\}^{z^{\prime}-N}$. 
Also, the prefactor $z^{\prime}!/N!\left(z^{\prime}-N\right)!$ stands for the number of combinations. 
This equation (\ref{A4}) gives the boundary condition for the stochastic time evolution Eq. (\ref{A5}). 
Considering the expectation as in the derivation of the McKendric equation (\ref{A6}), we have 
\begin{eqnarray}
\displaystyle \nonumber N_{t}\left(0,x\right)&:=&\displaystyle \sum_{N=1}^{\infty}NP_{t}^{\left(0,x\right)}\left(N\right)\\
\displaystyle \nonumber&=&\displaystyle \sum_{x^{\prime}\in\Omega}\int_{0}^{\infty}d\tau^{\prime}\,r\left(\tau^{\prime},x^{\prime}\right)N_{t}\left(\tau^{\prime},x^{\prime}\right)\\
\displaystyle \nonumber&&\times\biggl[\sum_{N=1}^{\infty}\sum_{z^{\prime}=N}^{\infty}\frac{Nz^{\prime}!}{N!\left(z^{\prime}-N\right)!}p\left(z^{\prime}|\tau^{\prime},x^{\prime}\right)\\
\nonumber&&\times\left(q\left(x;\tau^{\prime},x^{\prime}\right)\mathbb{T}\left(x|\tau^{\prime},x^{\prime}\right)\right)^{N}\\
&&\times\left(1-q\left(x;\tau^{\prime},x^{\prime}\right)\mathbb{T}\left(x|\tau^{\prime},x^{\prime}\right)\right)^{z^{\prime}-N}\biggr].
\end{eqnarray}
By changing the order of the summations $\Sigma_{N=1}^{\infty}$ and $\Sigma_{z^{\prime}=N}^{\infty}$, we get
\begin{eqnarray}
\displaystyle \nonumber&& N_{t}\left(0,x\right)=\sum_{x^{\prime}\in\Omega}\int_{0}^{\infty}d\tau^{\prime}\,r\left(\tau^{\prime},x^{\prime}\right)N_{t}\left(\tau^{\prime},x^{\prime}\right)\sum_{z^{\prime}=1}^{\infty}p\left(z^{\prime}|\tau^{\prime},x^{\prime}\right)\\
\displaystyle \nonumber&&\times\biggl[\sum_{N=1}^{z^{\prime}}\frac{z^{\prime}!}{\left(N-1\right)!\left(z^{\prime}-N\right)!}\left(q\left(x;\tau^{\prime},x^{\prime}\right)\mathbb{T}\left(x|\tau^{\prime},x^{\prime}\right)\right)^{N}\\
\nonumber&&\times\left(1-q\left(x;\tau^{\prime},x^{\prime}\right)\mathbb{T}\left(x|\tau^{\prime},x^{\prime}\right)\right)^{z^{\prime}-N}\biggr]
\end{eqnarray}
\begin{eqnarray}
\displaystyle \nonumber&=&\displaystyle \sum_{x^{\prime}\in\Omega}\int_{0}^{\infty}d\tau^{\prime}\,r\left(\tau^{\prime},x^{\prime}\right)N_{t}\left(\tau^{\prime},x^{\prime}\right)\\
\displaystyle \nonumber&&\times\sum_{z^{\prime}=1}^{\infty}z^{\prime}p\left(z^{\prime}|\tau^{\prime},x^{\prime}\right)q\left(x;\tau^{\prime},x^{\prime}\right)\mathbb{T}\left(x|\tau^{\prime},x^{\prime}\right)\\
\displaystyle \nonumber&&\times\biggl[\sum_{N=0}^{z^{\prime}-1}\frac{\left(z^{\prime}-1\right)!}{N!\left(\left(z^{\prime}-1\right)-N\right)!}\left(q\left(x;\tau^{\prime},x^{\prime}\right)\mathbb{T}\left(x|\tau^{\prime},x^{\prime}\right)\right)^{N}\\
&&\times\left(1-q\left(x;\tau^{\prime},x^{\prime}\right)\mathbb{T}\left(x|\tau^{\prime},x^{\prime}\right)\right)^{\left(z^{\prime}-1\right)-N}\biggr].
\end{eqnarray}
Finally, by using the binomial theorem: 
\begin{equation}
1=\displaystyle \sum_{N=0}^{z}\frac{z!}{N!\left(z-N\right)!}x^{N}\left(1-x\right)^{z-N},
\end{equation}
we obtain the boundary condition for the McKendric equation (\ref{A6}) as 
\begin{eqnarray}
\displaystyle \nonumber N_{t}\left(0,x\right)&=&\displaystyle \sum_{x^{\prime}\in\Omega}\int_{0}^{\infty}d\tau^{\prime}\,r\left(\tau^{\prime},x^{\prime}\right)N_{t}\left(\tau^{\prime},x^{\prime}\right)\\
\displaystyle \nonumber&&\times\sum_{z^{\prime}=1}^{\infty}z^{\prime}p\left(z^{\prime}|\tau^{\prime},x^{\prime}\right)q\left(x;\tau^{\prime},x^{\prime}\right)\mathbb{T}\left(x|\tau^{\prime},x^{\prime}\right)\\
\displaystyle \nonumber&=&\displaystyle \sum_{x^{\prime}\in\Omega}\int_{0}^{\infty}d\tau^{\prime}\,q\left(x;\tau^{\prime},x^{\prime}\right)\mathbb{T}\left(x|\tau^{\prime},x^{\prime}\right)\\
&&\times b\left(\tau^{\prime},x^{\prime}\right)r\left(\tau^{\prime},x^{\prime}\right)N_{t}\left(\tau^{\prime},x^{\prime}\right),
\end{eqnarray}
where we define the expected number of the newborn cells by $b\left(\tau^{\prime},x^{\prime}\right):=\Sigma_{z^{\prime}=1}^{\infty}z^{\prime}p\left(z^{\prime}|\tau^{\prime},x^{\prime}\right)$. 

\section{Appendix B}
In this appendix, we solve the McKendric equation (\ref{McPD}) by employing the eigenfunction method. 
The formal solution of Eq. (\ref{McPD}) is given by 
\begin{equation}
N_{t}\left(a,x\right)=e^{\hat{H}t}N_{0}\left(a,x\right),
\end{equation}
where $\hat{H}$ is the time evolution operator, Eq. (\ref{H}).
Therefore, by employing the eigenfunctions of $\hat{H}$, we can represent the solution as 
\begin{equation}
N_{t}\displaystyle \left(a,x\right)=\sum_{i=0}^{\infty}e^{\lambda_{i}t}C_{i}v_{i}\left(a,x\right),\label{21}
\end{equation}
where $\lambda_{i}$ and $v_{i}\left(a,x\right)$ denote the $\left(i+1\right)$th eigenvalue and the corresponding eigenfunction, respectively; 
especially, we determine the index $i$ so that $\lambda_{0}$ represents the largest eigenvalue, which means that ${\rm Re}\left[\lambda_{0}\right]\geq{\rm Re}\left[\lambda_{i}\right]$ for any $i$. 
Also, $\left\{C_{i}\right\}$ represent the expansion coefficients of the initial population $N_{0}\left(a,x\right)$: $N_{0}\left(a,x\right)=\Sigma_{i=0}^{\infty}C_{i}v_{i}\left(a,x\right)$. 
In this study, we assume that $\lambda_{0}$ is unique and real positive value, $\lambda_{0}>{\rm Re}\left[\lambda_{i}\right]$ ($i\neq 0$); 
that is, we deal with cases that the total size of the population is expanding in time evolution. 
Since the left hand side of Eq. (\ref{21}) is dominated by $v_{0}\left(a,x\right)$ as $ t\rightarrow\infty$, the fraction of the population converges to the unique stationary one $v_{0}\left(a,x\right)$ up to a normalizing constant. 
Furthermore, taking into account that $N_{t}^{\mathrm{t}\mathrm{o}\mathrm{t}}/N_{0}^{\mathrm{t}\mathrm{o}\mathrm{t}}\approx e^{\lambda_{0}t}$ for $ t\rightarrow\infty$, we find that the stationary population growth rate Eq. (\ref{SPGR}) is given by the largest eigenvalue, that is $\lambda=\lambda_{0}$. 
Accordingly, the calculation of the stationary population growth rate reduces to the eigenvalue problem of the time evolution operator $\hat{H}$ under the boundary condition Eq. (\ref{bMcPD}). 
To calculate the eigenvalues of $\hat{H}$, we consider the following characteristic equation: 
\begin{equation}
\lambda_{i}v_{i}\left(a,x\right)=\left[-\frac{\partial}{\partial a}-\left\{\gamma\left(a,x\right)+r\left(a,x\right)\right\}\right]v_{i}\left(a,x\right).\label{vCE}
\end{equation}
Then, the general solution of Eq. (\ref{vCE}) can be represented by 
\begin{equation}
v_{i}\left(a,x\right)=v_{i}\left(0,x\right)e^{-\lambda_{i}a}e^{-\int_{0}^{a}\left\{\gamma\left(t,x\right)+r\left(t,x\right)\right\}dt}.
\end{equation}
By using the boundary condition Eq. (\ref{bMcPD}), we obtain the constraint condition of $v_{i}\left(0,x\right)$ as 
\begin{eqnarray}
\displaystyle \nonumber&& v_{i}\left(0,x\right)=\sum_{x^{\prime}\in\Omega}\int_{0}^{\infty}d\tau^{\prime}\,q\left(x;\tau^{\prime},x^{\prime}\right)\mathbb{T}\left(x|\tau^{\prime},x^{\prime}\right)\\
\nonumber&&\times b\left(\tau^{\prime},x^{\prime}\right)r\left(\tau^{\prime},x^{\prime}\right)e^{-\lambda_{i}\tau^{\prime}}e^{-\int_{0}^{\tau^{\prime}}\left\{\gamma\left(t,x^{\prime}\right)+r\left(t,x^{\prime}\right)\right\}dt}v_{i}\left(0,x^{\prime}\right)\\
&&=\displaystyle \sum_{x^{\prime}\in\Omega}\int_{0}^{\infty}d\tau^{\prime}\,e^{k\left(x;\tau^{\prime},x^{\prime}\right)-\lambda_{i}\tau^{\prime}}Q\left(x;\tau^{\prime}|x^{\prime}\right)v_{i}\left(0,x^{\prime}\right),\label{vsceq}
\end{eqnarray}
where $k\left(x;\tau^{\prime},x^{\prime}\right)$ and $Q\left(x;\tau^{\prime}|x^{\prime}\right)$ are the growth kernel Eq. (\ref{defk}) and the semi-Markov kernel Eq. (\ref{defQ}). 
Thus, from Eq. (\ref{vsceq}), the eigenvalues and eigenfunctions are determined so that the self-consistent equation Eq. (\ref{vsceq}) is satisfied. 
This statement can be rephrased as follows. 
If we define a matrix 
\begin{equation}
M_{\alpha}\displaystyle \left(x|x^{\prime}\right):=\int_{0}^{\infty}d\tau^{\prime}\,e^{k\left(x;\tau^{\prime},x^{\prime}\right)-\alpha\tau^{\prime}}Q\left(x;\tau^{\prime}|x^{\prime}\right),\label{M}
\end{equation} 
the stationary population growth rate $\lambda$ is given by the largest $\alpha$ such that the eigenvalue of $M_{\alpha}\left(x|x^{\prime}\right)$ is unit. 
In addition, the stationary population with age $0,\ v_{0}\left(0,x\right)$, is calculated by the right eigenvector of $\left.M_{\alpha}\left(x|x^{\prime}\right)\right|_{\alpha=\lambda}$, the corresponding eigenvalue of which is unit. 

\section{Appendix C}
In this appendix, we derive Eq. (\ref{Nagai}), which is known as the many-to-one formula in the field of population genetics. 
The original proof of this formula is rigorously given in Refs. \cite{a6,a7}; however, we here demonstrate an alternative derivation that is familiar to physicists and mathematical biologists. 

Denote a tree of cell lineages during time interval $\left[0,T\right]$ by $\mathcal{T}_{T}$, which is a stochastic tree and its probability laws are given by the setup introduced in Sec. II and III. 
First, for simplicity, we consider cases that the lineage tree $\mathcal{T}_{T}$ is generated from a single root cell with age $0$, the mother cell of which has an inter-division interval $\tau_{0}$ and a type $x_{0}$; 
furthermore, for a mathematical implementation, we assume that the root cell has not yet undergone a type transition process, which means that calculation to obtain the lineage tree $\mathcal{T}_{T}$ starts with the type transition $\mathbb{T}\left(\cdot|\tau_{0},x_{0}\right)$ (see FIG. 7). 
\begin{figure}[h]
\begin{center}
\includegraphics*[height=7.5cm]{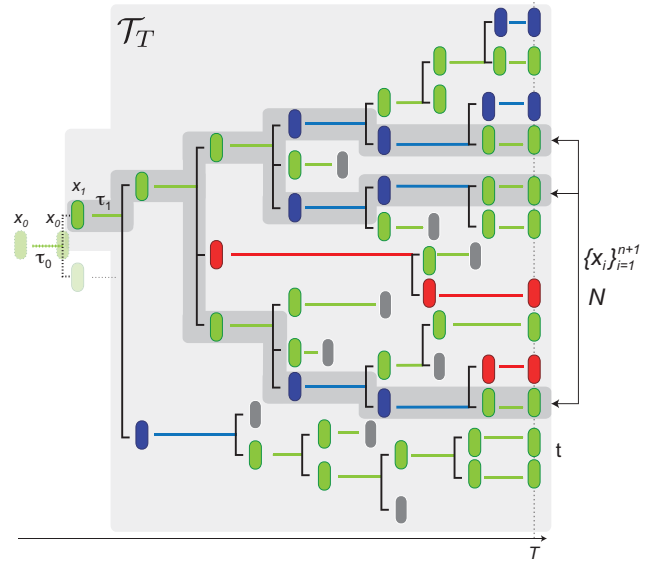}
\caption{A schematic illustration of the lineage tree $\mathcal{T}_{T}$. 
The root cell is one of the daughters generated from the mother cell with $\left(\tau_{0},x_{0}\right)$. 
The paths highlighted by the dark gray color represent surviving genealogical paths with history $\left\{\bm{x}_{i}\right\}_{i=1}^{n+1}$. 
The lineage tree $\mathcal{T}_{T}$ highlighted by the light gray color has $N$ surviving genealogical paths.
$P_{T}^{\left\{\bm{x}_{i}\right\}_{i=1}^{n+1}}\left(N|\tau_{0},x_{0}\right)$ represents the conditional probability that such a lineage tree $\mathcal{T}_{T}$ appears. }
\end{center}
\end{figure}
Under the above assumptions, we consider the conditional probability that the root cell generates a lineage tree $\mathcal{T}_{T}$ including $N$ surviving genealogical paths specified by $\left\{\bm{x}_{i}\right\}_{i=1}^{n+1}:=\left\{n;\bm{x}_{1},\bm{x}_{2},...,\bm{x}_{n},\bm{x}_{n+1}\right\}$. 
The vector $\bm{x}_{i}=\left(x_{i},\tau_{i},z_{i}\right)$ represents a type $x_{i}$, an age $\tau_{i}$ and the number of the daughter cells $z_{i}$ at the $i$th division event. 
$\bm{x}_{n+1}$ is specially defined as $\bm{x}_{n+1}=\left(x_{n+1},\tau_{n+1}\right)$. 
Also, the surviving genealogical path means the path penetrating the lineage tree $\mathcal{T}_{T}$ from the initial to the final time, i.e., cells have never undergone the death event on this path (see FIG. 7). 
We denote this conditional probability by $P_{T}^{\left\{\bm{x}_{i}\right\}_{i=1}^{n+1}}\left(N|\tau_{0},x_{0}\right)$, which is calculated by the following recurrence equation: 
\begin{eqnarray}
\nonumber&& P_{T}^{\left\{\bm{x}_{i}\right\}_{i=1}^{n+1}}\left(N|\tau_{0},x_{0}\right)=G\left(\bm{x}_{1}|\tau_{0},x_{0}\right)\\
\displaystyle \nonumber&&\times\sum_{\left\{\alpha_{l}\right\}:\Sigma_{l=1}^{z_{1}}\alpha_{l}=N}\left\{\prod_{l=1}^{z_{1}}P_{T-\tau_{1}}^{\left\{\bm{x}_{i}\right\}_{i=2}^{n+1}}\left(\alpha_{l}|\tau_{1},x_{1}\right)\right\}\\
&&+\delta_{N,0}\left(1-G\left(\bm{x}_{1}|\tau_{0},x_{0}\right)\right).\label{rec1}
\end{eqnarray}
Here, $G\left(\bm{x}|\tau^{\prime},x^{\prime}\right)$ denotes the conditional probability that the mother cell with type $x^{\prime}$ has divided at age $\tau^{\prime}$ and its daughter cell with $x$ survives for time $\tau$ and divides into $z$ cells; 
that is 
\begin{eqnarray}
\nonumber G\left(\bm{x}|\tau^{\prime},x^{\prime}\right)&:=&p\left(z|\tau,x\right)r\left(\tau,x\right)e^{-\int_{0}^{\tau}\left\{\gamma\left(t,x\right)+r\left(t,x\right)\right\}dt}\\
&&\times q\left(x|\tau^{\prime},x^{\prime}\right)\mathbb{T}\left(x|\tau^{\prime},x^{\prime}\right).\label{C2}
\end{eqnarray} 
Also, $P_{T-\tau_{1}}^{\left\{\bm{x}_{i}\right\}_{i=2}^{n+1}}\left(\alpha_{l}|\tau_{1},x_{1}\right)$ represents the conditional probability that a root cell, the mother cell of which has $\left(\tau_{1},x_{1}\right)$, generates a lineage tree $\mathcal{T}_{T-\tau_{1}}$ including $\alpha_{l}$ surviving genealogical paths specified by $\left\{\bm{x}_{i}\right\}_{i=2}^{n+1}:=\left\{n-1;\bm{x}_{2},...,\bm{x}_{n},\bm{x}_{n+1}\right\}$; 
furthermore, $P_{T-\tau_{1}}^{\left\{\bm{x}_{i}\right\}_{i=2}^{n+1}}\left(\alpha_{l}|\tau_{1},x_{1}\right)$ can be calculated by the same recurrence equation as Eq. (\ref{rec1}): 
\begin{eqnarray}
\nonumber&& P_{T-\tau_{1}}^{\left\{\bm{x}_{i}\right\}_{i=2}^{n+1}}\left(N|\tau_{1},x_{1}\right)=G\left(\bm{x}_{2}|\tau_{1},x_{1}\right)\\
\displaystyle \nonumber&&\times\sum_{\left\{\alpha_{l}\right\}:\Sigma_{l=1}^{z_{2}}\alpha_{l}=N}\left\{\prod_{l=1}^{z_{2}}P_{T-\Sigma_{j=1}^{2}\tau_{j}}^{\left\{\bm{x}_{i}\right\}_{i=3}^{n+1}}\left(\alpha_{l}|\tau_{2},x_{2}\right)\right\}\\
&&+\delta_{N,0}\left(1-G\left(\bm{x}_{2}|\tau_{1},x_{1}\right)\right).\label{rec2}
\end{eqnarray}
Accordingly, by iteratively employing the recurrence equation (\ref{rec1}), we obtain $P_{T}^{\left\{\bm{x}_{i}\right\}_{i=1}^{n+1}}\left(N|\tau_{0},x_{0}\right)$. 
The final recurrence is specially given by 
\begin{eqnarray}
\nonumber&& P_{T-\Sigma_{j=1}^{n}\tau_{j}}^{\left\{\bm{x}_{n+1}\right\}}\left(N|\tau_{n},x_{n}\right)=\delta_{N,1}F\left(\tau_{n+1},x_{n+1}|\tau_{n},x_{n}\right)\\
&&+\delta_{N,0}\left\{1-F\left(\tau_{n+1},x_{n+1}|\tau_{n},x_{n}\right)\right\},\label{recf}
\end{eqnarray}
where $F\left(\tau,x|\tau^{\prime},x^{\prime}\right)$ represents the conditional probability that the daughter cell, the mother cell of which has $\left(\tau^{\prime},x^{\prime}\right)$, switches its type to $x$ and survives for $\tau$; 
that is 
\begin{eqnarray}
\nonumber F\left(\tau,x|\tau^{\prime},x^{\prime}\right)&:=&e^{-\int_{0}^{\tau}\left\{\gamma\left(t,x\right)+r\left(t,x\right)\right\}dt}\\
&&\times q\left(x;\tau^{\prime},x^{\prime}\right)\mathbb{T}\left(x|\tau^{\prime},x^{\prime}\right).\label{C3}
\end{eqnarray}
For convenience, we rewrite the recurrence equations (\ref{rec1}) and (\ref{recf}) by using the moment generating function as 
\begin{eqnarray}
\displaystyle \nonumber&& m_{T}^{\left\{\bm{x}_{i}\right\}_{i=1}^{n+1}}\left(k|\tau_{0},x_{0}\right):=\sum_{N=0}^{\infty}e^{kN}P_{T}^{\left\{\bm{x}_{i}\right\}_{i=1}^{n+1}}\left(N|\tau_{0},x_{0}\right)\\
\nonumber&&=G\left(\bm{x}_{1}|\tau_{0},x_{0}\right)\left[m_{T-\tau_{1}}^{\left\{\bm{x}_{i}\right\}_{i=2}^{n+1}}\left(k|\tau_{0},x_{0}\right)\right]^{z_{1}}\\
&&+\left(1-G\left(\bm{x}_{1}|\tau_{0},x_{0}\right)\right),\label{mom}
\end{eqnarray}
and
\begin{eqnarray}
\nonumber&& m_{T-\Sigma_{j=1}^{n}\tau_{j}}^{\left\{\bm{x}_{n+1}\right\}}\left(k|\tau_{n},x_{n}\right)=e^{k}F\left(\tau_{n+1},x_{n+1}|\tau_{n},x_{n}\right)\\
&&+\left\{1-F\left(\tau_{n+1},x_{n+1}|\tau_{n},x_{n}\right)\right\}.
\end{eqnarray}
Since we are now interested only in the expected number of the genealogical paths, we iteratively use the differentiation of the moment recurrence Eq. (\ref{mom}) with respect to $k$ and obtain 
\begin{eqnarray}
\displaystyle \nonumber&& N_{T}\left[\left\{\bm{x}_{i}\right\}_{i=1}^{n+1}|\tau_{0},x_{0}\right]:=\sum_{N=1}^{\infty}NP_{T}^{\left\{\bm{x}_{i}\right\}_{i=1}^{n+1}}\left(N|\tau_{0},x_{0}\right)\\
\nonumber&&=F\left(\tau_{n+1},x_{n+1}|\tau_{n},x_{n}\right)z_{n}G\left(\bm{x}_{n}|\tau_{n-1},x_{n-1}\right)\times\cdots\\
&&\times z_{3}G\left(\bm{x}_{3}|\tau_{2},x_{2}\right)z_{2}G\left(\bm{x}_{2}|\tau_{1},x_{1}\right)z_{1}G\left(\bm{x}_{1}|\tau_{0},x_{0}\right).\label{C1}
\end{eqnarray}

Next, we evaluate the expected number of genealogical paths $\tilde{\chi}_{T}:=\left\{n;x_{1},\tau_{1},x_{2},\tau_{2},...,x_{n},\tau_{n},x_{n+1},\tau_{n+1}\right\}$. 
Note that the number of the daughter cells $\left\{z_{i}\right\}_{i=1}^{n}$ is not assigned in the path $\tilde{\chi}_{T}$. 
Therefore, from the summation of Eq.(\ref{C1}) with respect to $\left\{z_{i}\right\}_{i=1}^{n}$, we obtain 
\begin{eqnarray}
\displaystyle \nonumber&& N_{T}\left[\tilde{\chi}_{T}|\tau_{0},x_{0}\right]=\sum_{\left\{z_{i}\right\}_{i=1}^{n}}N_{T}\left[\left\{\bm{x}_{i}\right\}_{i=1}^{n+1}|\tau_{0},x_{0}\right]\\
\nonumber&&=F\left(\tau_{n+1},x_{n+1}|\tau_{n},x_{n}\right)\left[\prod_{i=1}^{n}\sum_{z_{i}}z_{i}G\left(\bm{x}_{i}|\tau_{i-1},x_{i-1}\right)\right].\\\label{C4}
\end{eqnarray}
By substituting Eqs. (\ref{C2}) and (\ref{C3}) into Eq. (\ref{C4}), we get 
\begin{eqnarray}
\nonumber N_{T}&&\left[\tilde{\chi}_{T}|\tau_{0},x_{0}\right]=e^{-\int_{0}^{\tau_{n+1}}\left\{\gamma\left(a,x_{n+1}\right)+r\left(a,x_{n+1}\right)\right\}da}\\
\displaystyle \nonumber&&\times\biggl[\prod_{i=1}^{n}q\left(x_{i+1};\tau_{i},x_{i}\right)\mathbb{T}\left(x_{i+1}|\tau_{i},x_{i}\right)\\
\nonumber&&\times b\left(\tau_{i},x_{i}\right)e^{-\int_{0}^{\tau_{i}}\gamma\left(a,x_{i}\right)da}\pi\left(\tau_{i}|x_{i}\right)\biggr]\\
&&\times q\left(x_{1}|\tau_{0},x_{0}\right)\mathbb{T}\left(x_{1}|\tau_{0},x_{0}\right),\label{C5}
\end{eqnarray}
where $b\left(\tau_{i},x_{i}\right):=\Sigma_{z_{i}=1}^{\infty}z_{i}p\left(z_{i}|\tau_{i},x_{i}\right)$, and $\pi\left(\tau_{i}|x_{i}\right)$ represents the distribution of the inter-division interval defined by Eq. (\ref{defpi}).

Finally, we change the root condition of the lineage tree $\mathcal{T}_{T}$. 
Although we have assumed that the lineage tree $\mathcal{T}_{T}$ is generated by the root cell, the mother of which has $\left(\tau_{0},x_{0}\right)$, we here replace this dependence on the mother with age $a_{0}$ and type $x_{1}$ of the root cell;
that is, we consider the case that the root cell of a lineage tree has age $a_{0}$ and type $x_{1}$. 
By using the survival probability $F\left(a_{0},x_{1}|\tau_{0},x_{0}\right)$ and Eq. (\ref{C5}), we obtain the expected number of genealogical paths for the lineage tree with the root $\left(a_{0},x_{1}\right)$ as
\begin{eqnarray}
\displaystyle \nonumber N_{T}\left[\tilde{\chi}_{T}|a_{0},x_{1}\right]&=&\displaystyle \delta\left(T-\left\{\sum_{i=1}^{n+1}\tau_{i}-a_{0}\right\}\right)\frac{N_{T}\left[\tilde{\chi}_{T}|\tau_{0},x_{0}\right]}{F\left(a_{0},x_{1}|\tau_{0},x_{0}\right)}\\
\nonumber&=&\delta\left(T-\left\{\sum_{i=1}^{n+1}\tau_{i}-a_{0}\right\}\right)\\
\nonumber&&\times e^{-\int_{0}^{\tau_{n+1}}\left\{\gamma\left(a,x_{n+1}\right)+r\left(a,x_{n+1}\right)\right\}da}\\
\displaystyle \nonumber&&\times\biggl[\prod_{i=1}^{n}q\left(x_{i+1};\tau_{i},x_{i}\right)\mathbb{T}\left(x_{i+1}|\tau_{i},x_{i}\right)\\
\nonumber&&\times b\left(\tau_{i},x_{i}\right)e^{-\int_{0}^{\tau_{i}}\gamma\left(a,x_{i}\right)da}\pi\left(\tau_{i}|x_{i}\right)\biggr]\\
&&\times e^{\int_{0}^{a_{0}}\left\{\gamma\left(a,x_{1}\right)+r\left(a,x_{1}\right)\right\}da},
\end{eqnarray}
where $\delta\left(T-\left\{\sum_{i=1}^{n+1}\tau_{i}-a_{0}\right\}\right)$ stands for the constraint for the surviving path. 
By considering the case with multiple roots, the population of which is distributed as $N_{0}\left(a_{0},x_{1}\right)$, we obtain Eq. (\ref{Nagai}). 

\section{Appendix D}
Here, we derive the explicit form of the  typical triplet over a retrospective history, $j_{B}^{*}\left(x;\tau^{\prime},x^{\prime}\right)$, and demonstrate how the stationary growth rate $\lambda$ is calculated by the variational principle, Eq. (\ref{VP}), under given growth kernel $k\left(x;\tau^{\prime},x^{\prime}\right)$ and semi-Markov kernel $Q\left(x;\tau^{\prime}|x^{\prime}\right)$.  
The typical triplet for the retrospective history, $j_{B}^{*}\left(x;\tau^{\prime},x^{\prime}\right)$, is given by 
\begin{eqnarray}
\nonumber&& j_{B}^{*}\left(x;\tau^{\prime},x^{\prime}\right)=\\
\displaystyle \nonumber&&\arg\max_{j}\left\{\sum_{x,x^{\prime}\in\Omega}\int_{0}^{\infty}d\tau^{\prime}\,k\left(x;\tau^{\prime},x^{\prime}\right)j\left(x;\tau^{\prime},x^{\prime}\right)-I_{F}\left[j\right]\right\},\\
\end{eqnarray}
where the maximization is taken under the constraints: 
\begin{eqnarray}
&&\displaystyle \sum_{x\in\Omega}\int_{0}^{\infty}d\tau^{\prime}\,j\left(x;\tau^{\prime},x^{\prime}\right)=\sum_{x\in\Omega}\int_{0}^{\infty}d\tau^{\prime}\,j\left(x^{\prime};\tau^{\prime},x\right),\label{constshift}\\
&&\displaystyle \sum_{x,x^{\prime}\in\Omega}\int_{0}^{\infty}d\tau^{\prime}\,\tau^{\prime}j\left(x;\tau^{\prime},x^{\prime}\right)=1,\label{constnor}
\end{eqnarray}
where the former represents the shift-invariant property and the latter is the normalization condition. 
By using the Lagrange multiplier method, we find that $j_{B}^{*}\left(x;\tau^{\prime},x^{\prime}\right)$ satisfies
\begin{equation}
0=k\displaystyle \left(x;\tau^{\prime},x^{\prime}\right)-\log\frac{j_{B}^{*}\left(x;\tau^{\prime},x^{\prime}\right)}{Q\left(x;\tau^{\prime}|x^{\prime}\right)g_{B}^{*}\left(x^{\prime}\right)}+\log\frac{\phi\left(x\right)}{\phi\left(x^{\prime}\right)}-\alpha\tau^{\prime},\label{lag0}
\end{equation}
where $\log\phi\left(x\right)$ ($\phi\left(x\right)>0$) and $\alpha$ are the Lagrange multipliers corresponding to the constraints (\ref{constshift}) and (\ref{constnor}), respectively. 
By taking average of the both sides in Eq. (\ref{lag0}) with respect to $j_{B}^{*}\left(x;\tau^{\prime},x^{\prime}\right)$, we get 
\begin{eqnarray}
\displaystyle \nonumber 0&=&\displaystyle \lambda+\sum_{x,x^{\prime}\in\Omega}\int_{0}^{\infty}d\tau^{\prime}\,j_{B}^{*}\left(x;\tau^{\prime},x^{\prime}\right)\log\frac{\phi\left(x\right)}{\phi\left(x^{\prime}\right)}\\
&&-\displaystyle \alpha\sum_{x,x^{\prime}\in\Omega}\int_{0}^{\infty}d\tau^{\prime}\,\tau^{\prime}j_{B}^{*}\left(x;\tau^{\prime},x^{\prime}\right),
\end{eqnarray}
where we use Eq. (\ref{SGRj}). 
In addition, by employing the constraints (\ref{constshift}) and (\ref{constnor}), we obtain $\lambda=\alpha$. 
This result represents that the calculation of the population growth rate $\lambda$ can be reduced to that of the Lagrange multiplier $\alpha$.  
By using  the constraints (\ref{constshift}) and (\ref{constnor}), we evaluate the Lagrange multipliers $\phi\left(x\right)$ and $\alpha$. 
First, we rewrite Eq. (\ref{lag0}) as 
\begin{equation}
j_{B}^{*}\displaystyle \left(x;\tau^{\prime},x^{\prime}\right)=\phi\left(x\right)e^{k\left(x;\tau^{\prime},x^{\prime}\right)-\alpha\tau^{\prime}}Q\left(x;\tau^{\prime}|x^{\prime}\right)\frac{g_{B}^{*}\left(x^{\prime}\right)}{\phi\left(x^{\prime}\right)}.\label{Dapp3}
\end{equation}
The integrations of Eq. (\ref{Dapp3}) with respect to $\left(x;\tau^{\prime}\right)$ and $\left(\tau^{\prime},x^{\prime}\right)$ lead respectively to 
\begin{eqnarray}
1&=&\displaystyle \sum_{x\in\Omega}\phi\left(x\right)M_{\alpha}\left(x|x^{\prime}\right)\frac{1}{\phi\left(x^{\prime}\right)},\label{Dapp1}\\
g_{B}^{*}\displaystyle \left(x\right)&=&\displaystyle \sum_{x^{\prime}\in\Omega}\phi\left(x\right)M_{\alpha}\left(x|x^{\prime}\right)\frac{g_{B}^{*}\left(x^{\prime}\right)}{\phi\left(x^{\prime}\right)},\label{Dapp2}
\end{eqnarray}
where $M_{\alpha}\left(x|x^{\prime}\right)$ is defined in Eq. (\ref{M}). 
From Eq. (\ref{Dapp1}), we find that $\alpha$ is determined so that $M_{\alpha}\left(x|x^{\prime}\right)$ has the unit eigenvalue, and $\phi\left(x\right)$ represents the left eigenvector corresponding to it. 
(In the following discussion, we write $\phi\left(x\right)$ by $\phi_{\alpha}\left(x\right)$ to clarify the dependence of $\alpha$. ) 
Moreover, since $M_{\alpha}\left(x|x^{\prime}\right)$ is primitive, the eigenvector corresponding to the largest eigenvalue of $M_{\alpha}\left(x|x^{\prime}\right)$ consists of real and positive components and all the other eigenvectors must have at least one non-real or negative component, owing to the Perron-Frobenius theorem. 
Taking $\phi_{\alpha}\left(x\right)>0$ into account, we find that the unit eigenvalue of $M_{\alpha}\left(x|x^{\prime}\right)$ must be the largest one. 
Thus, $\alpha$ is determined such that the largest eigenvalue of $M_{\alpha}\left(x|x^{\prime}\right)$ is unit. 
On the other hand, from Eq. (\ref{Dapp2}), we obtain 
\begin{equation}
g_{B}^{*}\left(x\right)=\phi_{\alpha}\left(x\right)\psi_{\alpha}\left(x\right),\label{Dapp4}
\end{equation}
where $\psi_{\alpha}\left(x\right)$ denotes the right eigenvector corresponding to the unit eigenvalue (i.e., the largest eigenvalue) of $M_{\alpha}\left(x|x^{\prime}\right)$. 
By substituting Eq. (\ref{Dapp4}) into Eq. (\ref{Dapp3}), we obtain the explicit form of $j_{B}^{*}\left(x;\tau^{\prime},x^{\prime}\right)$ as 
\begin{equation}
j_{B}^{*}\left(x;\tau^{\prime},x^{\prime}\right)=\phi_{\alpha}\left(x\right)e^{k\left(x;\tau^{\prime},x^{\prime}\right)-\alpha\tau^{\prime}}Q\left(x;\tau^{\prime}|x^{\prime}\right)\psi_{\alpha}\left(x^{\prime}\right).\label{expj}
\end{equation}
Here, we note that the normalization condition of $j_{B}^{*}\left(x;\tau^{\prime},x^{\prime}\right)$ has not yet been determined in Eq. (\ref{expj}) and the uniqueness of the Lagrange multiplier $\alpha$ has not yet been elucidated. 
The normalization condition can be given by constraint (\ref{constnor}), that is 
\begin{equation}
1=-\displaystyle \sum_{x,x^{\prime}\in\Omega}\phi_{\alpha}\left(x\right)\left\{\frac{d}{d\alpha}M_{\alpha}\left(x|x^{\prime}\right)\right\}\psi_{\alpha}\left(x^{\prime}\right),\label{Dapp5}
\end{equation}
where we substitute Eq. (\ref{expj}) into Eq. (\ref{constnor}). 
Furthermore, by using Eq. (\ref{Dapp5}), we can clarify uniqueness of $\alpha$ as follows.
Denote $\xi_{\alpha}$ by the largest eigenvalue of $M_{\alpha}\left(x|x^{\prime}\right)$. 
Since the differentiation of $\xi_{\alpha}$ satisfies \cite{a1} 
\begin{equation}
\displaystyle \frac{d\xi_{\alpha}}{d\alpha}=\frac{\sum_{x,x^{\prime}\in\Omega}\phi_{\alpha}\left(x\right)\left\{\frac{d}{d\alpha}M_{\alpha}\left(x|x^{\prime}\right)\right\}\psi_{\alpha}\left(x^{\prime}\right)}{\sum_{x\in\Omega}\phi_{\alpha}\left(x\right)\psi_{\alpha}\left(x\right)}<0,
\end{equation}
the largest eigenvalue $\xi_{\alpha}$ is monotonically decreasing with respect to $\alpha$. 
Since Eq. (\ref{Dapp1}) leads to $\xi_{\alpha}=1$, we find that the Lagrange multiplier $\alpha$ is uniquely determined. 

Finally, we summarize the discussion of this appendix. 
The stationary population growth rate $\lambda$ is uniquely determined such that the largest eigenvalue of $M_{\lambda}\left(x|x^{\prime}\right)$ is unit. 
Also, by using the left and right eigenvectors corresponding to the largest (=unit) eigenvalue of $M_{\lambda}\left(x|x^{\prime}\right)$, the explicit form of the  typical triplet for the retrospective history can be represented as 
\begin{equation}
j_{B}^{*}\left(x;\tau^{\prime},x^{\prime}\right)=\phi_{\lambda}\left(x\right)e^{k\left(x;\tau^{\prime},x^{\prime}\right)-\lambda\tau^{\prime}}Q\left(x;\tau^{\prime}|x^{\prime}\right)\psi_{\lambda}\left(x^{\prime}\right).
\end{equation}
Since the above statement is corresponding to the result in the partial-differential-equation approach introduced in Appendix B, we find that $\psi_{\lambda}\left(x\right)$ and $\phi_{\lambda}\left(x\right)$ correspond to $v_{0}\left(0,x\right)$ and $u_{0}\left(0,x\right)$, respectively. 


\end{document}